\documentclass[letterpaper, 10 pt, journal]{IEEEtran}
\IEEEoverridecommandlockouts                          
\usepackage{graphics} 
\usepackage{epsfig}
\usepackage{times}
\usepackage{amsmath}
\usepackage{amssymb}
\usepackage{threeparttable}
\usepackage{booktabs}
\usepackage{soul}
\usepackage{subfigure}
\usepackage{xcolor}
\usepackage[hidelinks]{hyperref}
\usepackage{algorithm}
\usepackage{algorithmic,comment}
\usepackage{tcolorbox}
\usepackage{multirow}
\usepackage{dblfloatfix,cite}
\usepackage[numbers,sort&compress]{natbib}
\usepackage{tcolorbox}
\usepackage{soul,xcolor}
\tcbuselibrary{breakable}
\usepackage{tabularx}
\usepackage{booktabs}
\usepackage{caption}
\usepackage{float}
\usepackage{url,amsmath}
\usepackage{amssymb}
\usepackage{tabularray}
\usepackage{multirow}
\UseTblrLibrary{booktabs}
\usepackage{array}
\usepackage[switch]{lineno}

\usepackage[inkscapelatex=false]{svg}
\setstcolor{red}
\tcbuselibrary{skins}
\newcolumntype{C}{>{\centering\arraybackslash}X} 
\def\tsc#1{\csdef{#1}{\textsc{\lowercase{#1}}\xspace}}
\tsc{WGM}
\tsc{QE}
\tsc{EP}
\tsc{PMS}
\tsc{BEC}
\tsc{DE}

\title{\LARGE \bf
Transforming Remanufacturing Automation with Large Language Models: A Forward-Looking Analysis with Case Studies
}

\author{Chang Liu$^{1}$, Sara Behdad$^{2}$, Prabhakar Pagilla$^{1}$, Xiao Liang$^{3,\dagger}$, and Minghui Zheng$^{1,\dagger}$
\thanks{This work was partially supported by the U.S. National Science Foundation under Grant No. 2527316, No. 2422826, No. 2324950, and No. 2026276. Portions of this research were conducted with the advanced computing resources provided by Texas A\&M High Performance Research Computing.}
\thanks{$^{1}$ Chang Liu, Prabhakar Pagilla and Minghui Zheng are with the J. Mike Walker '66 Department of Mechanical Engineering, Texas A\&M University, College Station, TX 77843, USA. {\tt\small Emails: {changliu.chris, ppagilla, mhzheng}@tamu.edu.}}
\thanks{$^{2}$ Sara Behdad is with the Engineering School of Sustainable
Infrastructure \& Environment, University of Florida, Gainesville, Florida 32611, USA. {\tt\small Email: sara.behdad@essie.ufl.edu.}}
\thanks{$^{3}$ Xiao Liang is with the Zachry Department of Civil and Environmental Engineering, Texas A\&M University, College Station, TX 77843, USA. {\tt\small Email: xliang@tamu.edu.}}
\thanks{$^\dagger$ Corresponding Authors.}}

\begin{document}

\maketitle
\thispagestyle{empty}
\pagestyle{empty}

\begin{abstract}

With growing concerns about resource scarcity and environmental degradation, remanufacturing of end-of-life (EoL) products within the circular economy is attracting increasing attention. Remanufacturing can preserve most of the original manufacturing value and materials while transforming EoL products into like-new condition. However, the variability and uncertainty of EoL products make remanufacturing highly dependent on human expertise. Although recent advances in robotics and artificial intelligence have improved automation in isolated remanufacturing tasks, existing methods are often task-specific and struggle to generalize to heterogeneous EoL conditions. Recently, large language models (LLMs) have demonstrated remarkable capabilities in learning from massive, unstructured datasets, generating expert-level output across various tasks, and communicating with humans in natural language for interpretation. These advantages can align closely with the complex demands of remanufacturing, thereby mitigating the reliance on specialized expertise. However, their roles and research progress in this domain remain underexplored. In this paper, we present a forward-looking review and analysis of the role of LLMs in remanufacturing automation, grounded in a brief critical review of existing LLM-related studies relevant to remanufacturing. Building on this foundation, we introduce ReManGPT as a conceptual framework and use three representative case studies to illustrate selected modules of the framework in practical remanufacturing scenarios. We also analyze three representative remanufacturing applications, electric vehicle batteries, electronic waste, and electric motors, to illustrate how the proposed framework could address their domain-specific challenges. Finally, we discuss the current barriers to deploying this framework in practice and outline future research directions, including LLM-assisted human operation and language–action models for robotic automation.

\end{abstract}

\section{Introduction}

\begin{figure}[t]
	\centering
	\includegraphics[width=0.48\textwidth]{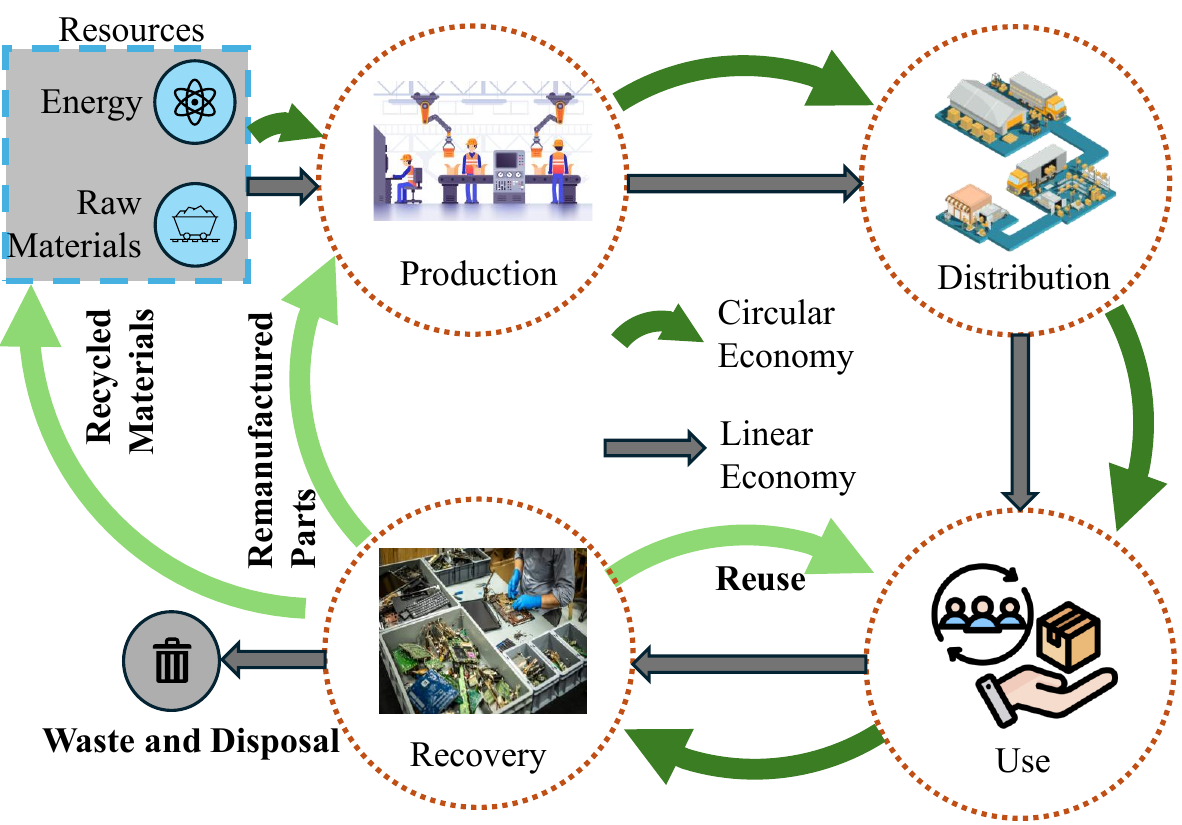}
        \vspace{3pt}
        \caption{The comparison between the linear and circular economies. Unlike the linear economy, which ends with waste and disposal, the circular economy closes the loop through reuse, remanufacturing, and recycling. Remanufacturing is highlighted because it preserves product value.
        }
	\label{circular}
    \vspace{-0.2in}
\end{figure}

The growing concern about natural resource shortages and environmental pollution problems has increased the need to find sustainable solutions in both the economic and manufacturing sectors \cite{geissdoerfer2017circular}. The circular economy is a viable solution to address these concerns and has attracted worldwide attention because it provides an alternative solution to the current linear economy of  ``Take, Make, Use, and Dispose'' \cite{frosch1989strategies}. It aims to minimize resource consumption and waste generation by closing energy and material loops \cite{despeisse2017unlocking}. As shown in Fig.~\ref{circular}, linear and circular economies differ in how materials and value are created, retained, and recovered.
Within the circular economy, end-of-life (EoL) products are processed primarily in three ways: reuse, recycling, and remanufacturing \cite{modak2023review}. Reuse typically refers to the use of a product within its first lifecycle for the same or a different purpose without making significant changes to the product. While it retains most of the original value of the EoL product, it cannot guarantee quality, resulting in low customer acceptance rates and profits. Recycling converts the EoL product components into raw materials for new production. However, eliminating the manufacturing value of products through the recycling process is considered less sustainable from both economic and environmental perspectives \cite{singhal2020remanufacturing, hunka2021determinants}.

\begin{figure*}[htp!]
	\centering
	\includegraphics[width=0.95\textwidth]{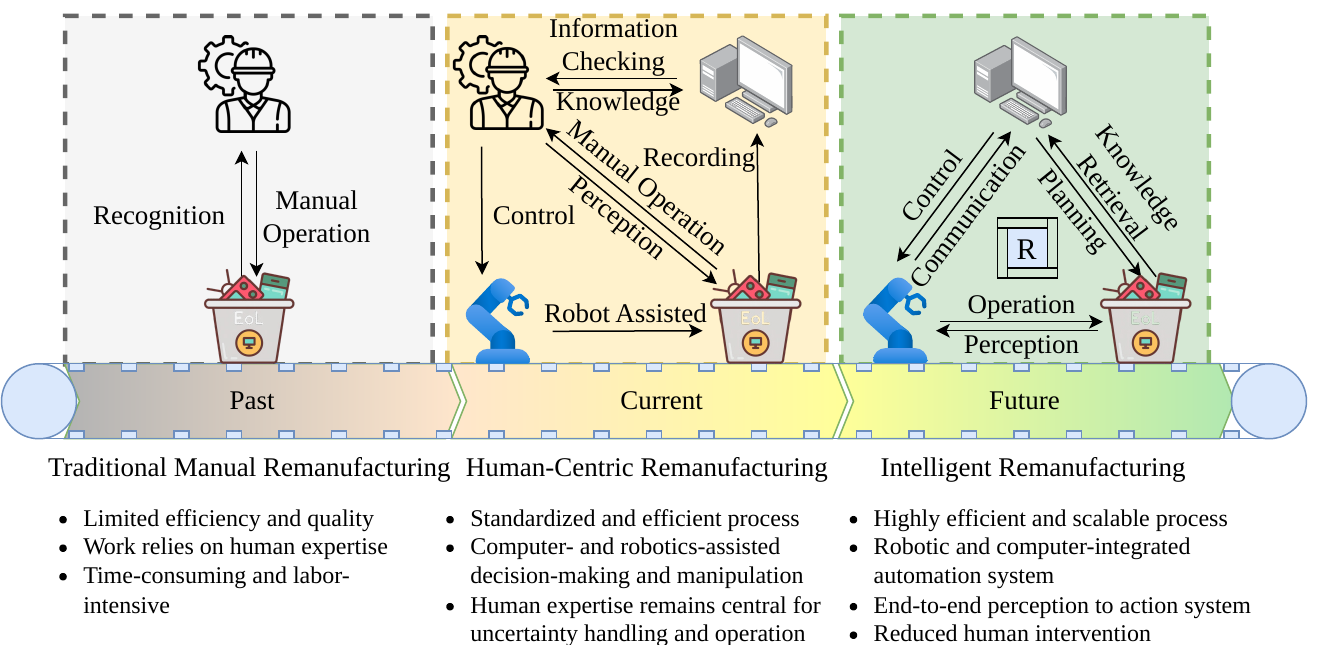}
        \vspace{3pt}
        \caption{Evolution of remanufacturing automation from traditional manual operation to human-centric automation and future intelligent remanufacturing. The figure highlights the shift from human-dependent recognition and operation to computer- and robot-assisted processes, and further toward ReManGPT-supported integration of perception, communication, planning, knowledge retrieval, and operation.
        }
	\label{fig:RMEvo}
\end{figure*}

Remanufacturing is a technique-intensive process of inspecting and sorting EoL products, disassembling the products, repairing, replacing, and recovering valuable components, and reassembling components into products with new life \cite{khan2022effective, chirumalla2023second}. Remanufacturing gives new life to EoL products, with quality and warranties equal to or better than those of new products \cite{king2006reducing}. Driven by global environmental regulations and increasing social expectations, governments are actively promoting remanufacturing-related policies \cite{goodall2014review, chong2024sustainable, zhang2024fairness}. In response to these policies, industries should reformulate their production strategies to incorporate remanufacturing, achieving both sustainability and economic benefits. In practice, lifecycle assessment (LCA) is frequently employed to evaluate their environmental impacts, while market analyses are used to identify potential economic opportunities \cite{zeng2024does, gunasekara2021remanufacture, jiang2016reliability, fadeyi2022instilling, fofou2021review}. Additionally, remanufacturing provides substantial environmental benefits, saving up to 85\% of energy and 70\% of materials compared to new production \cite{jiang2016reliability}. It also has significant economic value, with remanufactured products costing up to 40\% less while maintaining around 20\% profit margins for companies \cite{ilgin2012remanufacturing}. By reducing waste and giving products new life, remanufacturing has become essential for industries and regions facing material constraints. It also contributes to building a more resource-efficient global circular economy \cite{hazen2017remanufacturing, russell2023value, wu2025select, tsao2024remanufacturing}.

At the same time, the growing scarcity of critical materials, particularly rare earth elements (REEs), has attracted growing attention from industries that rely heavily on these resources. Because REE production is geographically concentrated, supply chains remain vulnerable to disruption and geopolitical constraints. Therefore, remanufacturing is increasingly recognized as an environmental and economic opportunity, as well as a strategic approach for mitigating material supply risks  \cite{maani2024disassembly, wang2024regional, koese2025dynamics}. Remanufacturing has been widely adopted in various industrial sectors, including aerospace, automotive, and electrical equipment. In the US, nearly 90\% of automobile replacement parts are remanufactured \cite{matsumoto2016trends}. The global market for machinery remanufacturing is valued at \$421.5 billion in 2025 and is expected to exceed \$2.5 trillion by 2034 \cite{USDAnalytics2025}. Several applications have emerged due to their high remanufacturing potential, rapidly growing volumes, and firms' reliance on REEs; key examples include energy storage systems, electric motors, and electronic waste. Although these application domains differ in scale, value, and operational uncertainty, they share a heavy reliance on expert judgment and limited levels of automation, making them well-suited for examining advanced remanufacturing intelligence.

Collected EoL products are typically under heterogeneous conditions and in various structures, requiring real-time decision-making, dynamic planning, and knowledge retrieval at each stage. Therefore, remanufacturing operations have relied heavily on human expertise. Traditionally, remanufacturing has been a manual process that is labor-intensive, time-consuming, and inefficient. Human workers performing remanufacturing learned from other experts and retrieved knowledge from unstructured documents; this requires a steep learning curve and can be challenging for adaptation to new products and techniques. These factors may limit the quality of the output and reduce the overall efficiency.

With recent advances in machine learning methods and robotics, industries have evolved from manual remanufacturing to a human-centric remanufacturing environment \cite{guo2024multi, sierra2024diagnosing, nwankpa2021achieving, caterino2025enhancing, rizova2020systematic}. These techniques have been employed to assist decision-making, planning, and process optimization. Human workers can identify products and check related records to gain knowledge of diagnoses and standardize operations. In this process, digital methods can be used to classify returned products, optimize disassembly sequences, and extract insights from sensor data or maintenance records. They offer improved accuracy and scalability over traditional manual approaches. Using the planning and decision-making results from these digital methods, humans can control the robot to perform some repetitive tasks in disassembly. At the same time, robots can collaborate with humans or automate selected repetitive tasks, reducing human workload and improving efficiency. However, robotic performance remains primitive in real-world industrial practice and is highly dependent on rigid planning results. This creates a technical barrier that prevents frontline experts without programming knowledge from adapting plans or controlling the robot effectively.

In addition, the methods that support decision-making, planning, and robot control are usually designed as fragmented solutions for specific tasks and rely on structured, high-quality, and labeled datasets. This prerequisite is fundamentally incompatible with the raw, fragmented information typically collected on remanufacturing lines, which requires substantial time and labor to process before it can be used. These models also require substantial feature engineering, domain-specific adjustment, and transfer learning to maintain performance in different scenarios. However, they generate outputs without providing clear explanations. This creates a cognitive gap, where human operators cannot understand the rationale behind decisions, ultimately undermining their trust in and reliance on these models.

Moreover, they lack the flexibility to generalize across different conditions or adapt to dynamic and uncertain decision environments \cite{kurilova2018remanufacturing, ngu2020review, matsumoto2016trends, psarommatis2025product}. These limitations still constrain humans within the remanufacturing loop as the primary decision-makers, planners, supervisors, and operators. This situation does not align with the escalating requirements of remanufacturing EoL products and the increasing labor shortage. 
Future progress depends on reducing the reliance on experienced workers for knowledge retrieval and decision-making. Remanufacturing systems must also support real-time planning and execution under uncertainty, while keeping their processes and outputs understandable and adjustable for operators. These conditions are essential for advancing remanufacturing toward higher levels of automation.
\begin{figure*}[htp!]
	\centering
	\includegraphics[width=0.9\textwidth]{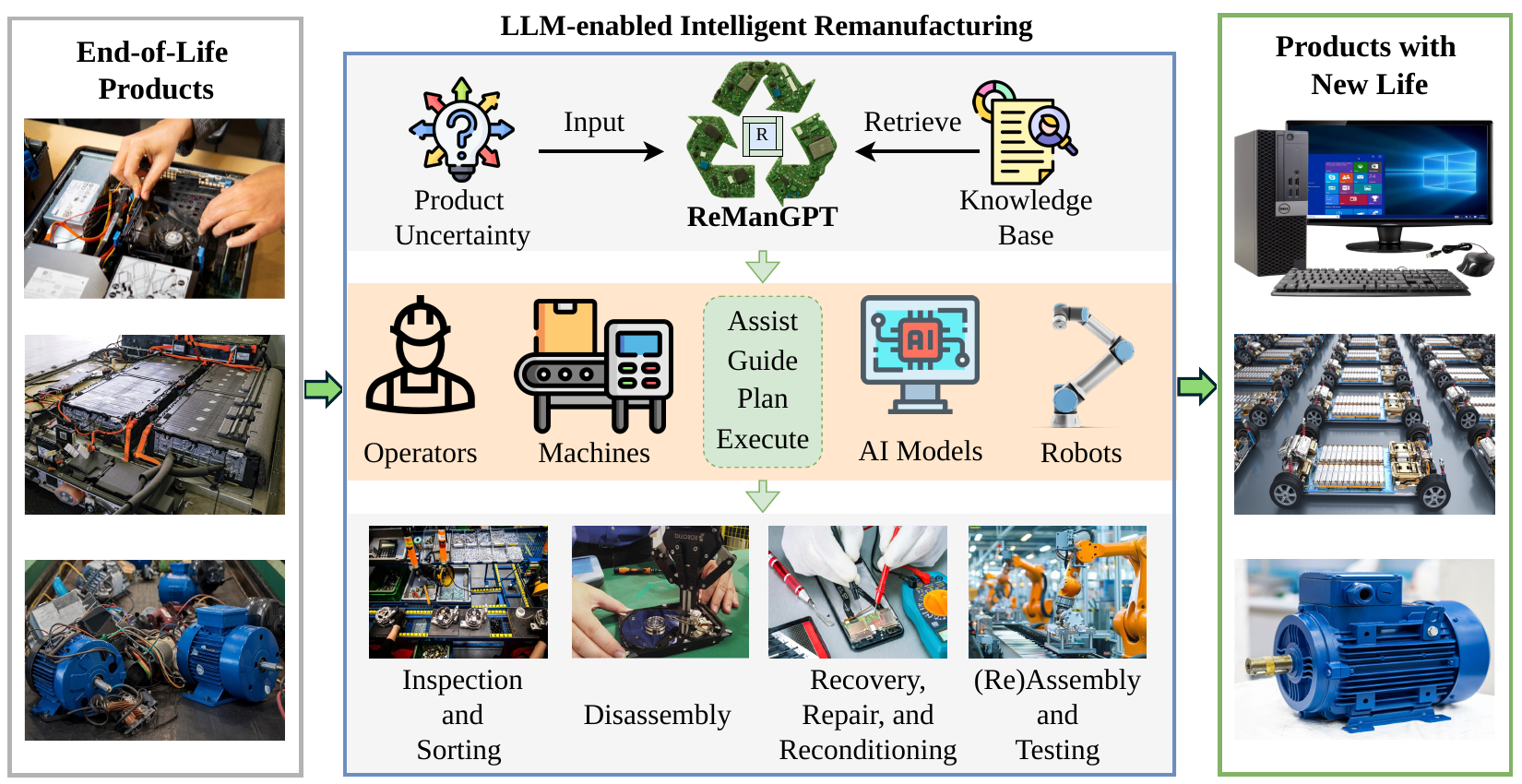}
        \vspace{3pt}
        \caption{Motivation for introducing ReManGPT in LLM-enabled intelligent remanufacturing. The left side shows heterogeneous EoL products with uncertain conditions and complex processing requirements. The middle part positions ReManGPT as a possible LLM-enabled framework that integrates product uncertainty, reusable knowledge, operators, machines, AI models, and robots. The bottom part maps this support to major remanufacturing stages, including inspection and sorting, disassembly, recovery, repair and reconditioning, and reassembly and testing. The right side shows the intended outcome of transforming EoL products into products with new life.}
	\label{fig:Motivation}
\end{figure*}

Recently, the rapid development of generative artificial intelligence, particularly Large Language Models (LLMs), has attracted widespread attention and created new possibilities for addressing these needs. Built upon transformer architectures \cite{vaswani2017attention} and pre-trained on massive corpora, LLMs have demonstrated emergent capabilities in natural language understanding and generation, pattern extraction from unstructured data, and few-shot adaptation to diverse tasks \cite{chen2023unleashing}. LLMs can synthesize domain-specific knowledge, interpret contextual information, and generate relevant, actionable outputs in natural language. These capabilities position LLMs as a promising tool for addressing complex multi-level planning and decision-making challenges in remanufacturing \cite{wang2024waste,clemm2024towards}. Several LLM capabilities are particularly relevant to remanufacturing, including fine-tuning, prompt engineering, augmentation through external knowledge and tools, and multi-modal modeling. These capabilities support domain-specific adaptation, instruction following, evidence integration, and the processing of non-textual information.
LLMs have been applied in various domains, such as finance \cite{jeong2024fine}, medicine \cite{thirunavukarasu2023large}, chemistry \cite{jablonka2024leveraging}, and industry \cite{chen2025integrating}.

In the current human-centric remanufacturing domain, LLMs can serve as intelligent digital assistants with domain expertise. Instead of learning from senior experts and searching through a massive set of documents for solutions, operators could communicate with LLM-based systems in natural language to retrieve context-aware instructions and diagnose malfunctioning parts. This approach also actively involves frontline operators in the decision-making loop. They can ask the model to explain and, based on their findings, subsequently modify or correct the plan. These operations and interactions can be logged by the system for later retrieval. LLMs can understand human requirements in natural language and convert them into commands that machines or robots can execute, enabling operators to control these systems without programming proficiency. These capabilities improve operational efficiency and reduce the dependence on expertise, reducing technical barriers within the current human-centric remanufacturing loop.

Looking ahead to the next generation of intelligent remanufacturing, LLMs have the potential to function as the cognitive core that supports higher levels of automation and broader generalization \cite{ding2026ccm}. Fig.~\ref{fig:RMEvo} illustrates the expected evolution of remanufacturing from traditional manual operations to human-centric automation and, ultimately, to intelligent, LLM-enabled remanufacturing systems.
Within such an intelligent remanufacturing framework, LLM-based agents can support task-oriented functions across different stages of the remanufacturing process. At early stages, these agents can combine multi-modal perception data about EoL products to support decision-making and initial task planning. During the operation, the model can integrate real-time observations and retrieve relevant domain knowledge to review and revise plans and decisions to handle the uncertainty of EoL components. These updated plans and decisions can then be translated into executable instructions or commands for robots and automated systems, closing the perception–to–action loop and enabling adaptive automation with reduced human intervention.

Although the rapid development of LLMs has demonstrated their success and potential in robotics, logistics, and manufacturing, their use in remanufacturing remains fragmented and underexplored. Existing studies typically address specific problems, such as disassembly sequence planning, component detection, and multi-modal data processing, while many other stages in the remanufacturing workflow have received limited attention. At the same time, this domain lacks a critical review that connects key LLM capabilities, including reasoning, planning, and natural language processing, to the practical challenges faced by human workers, machines, AI systems, and robotic platforms across remanufacturing operations. In addition, there is no forward-looking analysis that examines the potential impacts of LLMs on major remanufacturing applications. These gaps together motivate the work presented in this paper.

Motivated by the aforementioned observations, this paper presents a forward-looking review and analysis of the role of LLMs in remanufacturing automation. This analysis is grounded in a brief critical review of existing LLM-related studies on remanufacturing and is further developed using the ReManGPT framework and three representative case studies. ReManGPT is introduced as a conceptual framework for supporting knowledge retrieval, reasoning, planning, and execution across major remanufacturing stages under uncertain product conditions. The framework is proposed for higher-level human-centric remanufacturing and for the transition toward future intelligent remanufacturing. Fig.~\ref{fig:Motivation} further illustrates the motivation for introducing ReManGPT in the future intelligent remanufacturing stage shown in Fig.~\ref{fig:RMEvo}. It links uncertain EoL product conditions and fragmented process knowledge with ReManGPT's role in coordinating human operators, machines, AI models, and robots across major remanufacturing operations. The remainder of this paper is structured as follows. Section 2 provides a brief critical review of existing LLM-related studies across major stages of remanufacturing and identifies the main gaps in the current literature. Section 3 introduces the ReManGPT framework and presents three representative case studies to illustrate selected framework modules in practical remanufacturing scenarios. Section 4 discusses the broader implications for key remanufacturing applications. Section 5 discusses current limitations and challenges. Section 6 outlines future research directions. Section 7 concludes the paper.

\begin{figure*}[htp]
	\centering
	\includegraphics[width=0.9\textwidth]{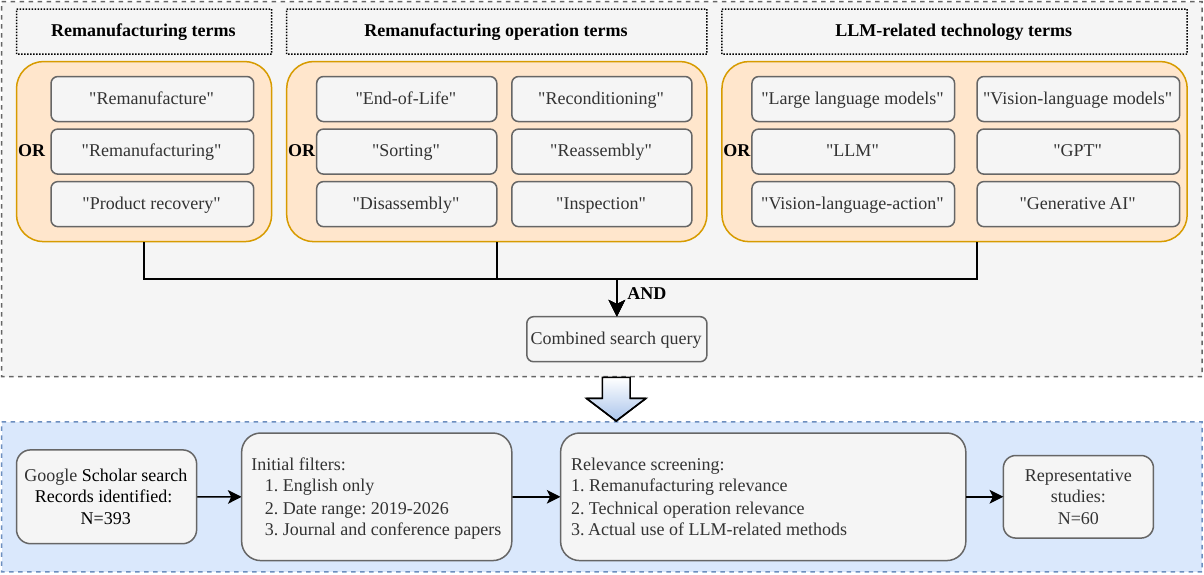}
        \vspace{3pt}
        \caption{Search and screening process for identifying representative LLM-related remanufacturing studies used in the brief critical review.
        }
	\label{fig:SearchMethod}
\end{figure*}

\section{Existing LLM-related Work in Remanufacturing}

This section provides a brief critical review of existing LLM-related studies across the major stages of remanufacturing. It begins with LLM applications in the major stages of remanufacturing, including inspection and sorting, disassembly, recovery, repair, and reconditioning, and reassembly with testing. It then turns to design for remanufacturing and process planning, which address remanufacturing at a higher level. Design for remanufacturing integrates remanufacturing and product lifecycle considerations into the product design phase, while process planning improves the organization of the overall remanufacturing process. The final part considers studies that reduce the overall reliance on human expertise and support the move toward higher levels of automation across stages or at the system level. This organization is essential because the ReManGPT framework is designed to coordinate knowledge retrieval, decision-making, planning, execution, and record transfer across the same remanufacturing stages.

To support this forward-looking analysis, we conducted a brief but structured search for LLM-related studies in remanufacturing. The search and screening process is summarized in Fig.~\ref{fig:SearchMethod}. We formulated search queries using three concept groups: remanufacturing terms, remanufacturing-operation terms, and LLM-related technology terms. Keywords within each group were combined using the ``OR'' operator. The three concept groups were then linked using the ``AND'' operator so that the retrieved studies addressed all core aspects relevant to this review. Because explicit studies at the intersection of LLMs and remanufacturing remain limited in conventional databases, Google Scholar was used as the primary search engine. The search returned 393 records, which were then filtered by language, publication type, and date range and screened for technical relevance and actual use of LLM-related methods. The final set included 60 representative studies for the critical review and analysis.

\subsection{LLMs in Inspection and Sorting}

Inspection and sorting are central to managing the uncertainty in collected EoL products. They help evaluate conditions, identify defects and damages, assess remanufacturing value, and route products or components to the different operations. Inspection appears at different points in remanufacturing. Before and after disassembly, it supports sorting and process planning. After reassembly, it serves as final quality control.

Current practice still depends heavily on expert-level manual inspection. This approach is time-consuming and susceptible to human error, such as missed defects \cite{cemenska2017afp}. Such errors can compromise downstream remanufacturing operations and reduce the value of the final product. To improve efficiency, existing studies have introduced semi-automated systems based on sensor data, computer vision, and machine learning methods \cite{saiz2021inspection,zhang2019remanufacturing,schluter2021ai, selvakanmani2024optimizing, khan2021vision}. These methods can support damage detection and condition assessment, with the results used for sorting, downstream decision-making, and process planning. However, most of them are limited to specific tasks and predefined settings. Under uncertain EoL product conditions, inspection cannot be fully standardized in advance. Reliable assessment requires integrating records, perception, and test results, while sorting depends on interpreted inspection results rather than on perception outputs alone. Automated inspection and sorting, therefore, remain a process-level challenge, requiring the integration of perception, assessment, decision-making, and precise non-destructive robotic execution \cite{kaiser2022concept, nwankpa2021design}.

LLM-based systems could support inspection and sorting by transforming process knowledge and multi-modal inputs into operational guidance. They could provide step-by-step inspection procedures under different product conditions, reducing the need for workers to search through scattered instructions during operation. Uncertain cases could be addressed through natural-language, text- or speech-based interactions, and the systems can output explanations and task-level guidance that reduce reliance on senior workers. Such systems could also record observations, test results, intermediate judgments, and final sorting decisions for subsequent processes, thereby reducing manual documentation. At a higher level of automation, they may further translate complex inspection procedures into executable robotic steps for fine-grained and non-destructive inspection. Tasneem et al. propose utilizing LLMs in the human–robot collaborative (HRC) inspection operation \cite{tasneem2026human}. The operator can communicate with the LLMs directly in natural language through the speech-to-text model, allowing the LLMs to generate executable code for the robot to complete the inspection.

\subsection{LLMs in Disassembly}

Disassembly is a vital step in the remanufacturing of EoL products, as its cost and efficiency largely influence the success of the overall process \cite{he2024disassembly}. In practice, disassembly often involves considerable uncertainty and complexity because EoL products vary widely in condition. Human workers, therefore, remain heavily involved, given their ability to adapt to such variability. However, growing labor shortages and the rising volume of EoL products are making this reliance on manual disassembly increasingly difficult to sustain and economically unattractive \cite{wu2026empowering, liu2026batch, liu2026raise}.

Disassembly sequence planning (DSP) focuses on finding a detailed disassembly plan to remove specific components from the EoL product.
It determines what should be disassembled, in which mode, and how the process may be optimized with respect to cost and time. Existing research on DSP usually relies on predefined product models, from which precedence relations and constraint information are derived before sequence optimization is performed. Related problems such as task allocation \cite{lee2022task}, task planning \cite{lou2024human,gao2024partially}, and disassembly line balancing \cite{duan2026enhancing, zhang2026green, lou2024personalized} are then addressed on the same basis. In practice, however, these models are often built by human experts from original CAD data. Nevertheless, sequence planning and optimization results are typically computed offline, which requires considerable computational resources. Additionally, uncertainty in disassembly is also commonly reduced to predefined parameters \cite{hsu2016fuzzy}. This leaves a clear gap between existing DSP methods and real EoL disassembly practice, where the required information is often incomplete or unavailable \cite{klein2026study}. To narrow this gap, recent studies have begun to move beyond offline DSP toward more adaptive disassembly systems. One practical research direction is HRC disassembly \cite{ding2026review}, in which humans can address uncertainties in EoL products beyond predefined plans and take over operations that are still difficult for robots to perform reliably. Although HRC offers a practical response to uncertainty in disassembly, it does not fundamentally remove reliance on human intervention. System perception, therefore, remains necessary for generating a feasible disassembly sequence and performing decision-making from current sensory input and available knowledge, rather than relying mainly on humans.
Several studies have explored distributed and multi-modal solutions to address this problem. For example, Ferrari et al. proposed a multi-agent robotic software cell for complex EV battery disassembly \cite{ferrari2025distributed}. Liu et al. \cite{liu2026robotic} introduced a digital twin–based robotic disassembly framework integrated with deep reinforcement learning to dynamically perform DSP in the presence of uncertainties from components. Zhang et al. proposed a multi-objective HRC disassembly line balancing framework with a hybrid optimization model and a Q-learning algorithm for the destructive disassembly process \cite{zhang2026green}. Lv et al. presented a multi-cognitive robotic agent system that integrates scene graphs, knowledge graphs, and an attention-based agent to dynamically enhance the HRC disassembly decision-making process \cite{lv2026graph}.

Recent studies show that LLM-related methods have been integrated into the disassembly workflows in several distinct but connected directions. Some work remains focused on planning-related problems, introducing LLMs into DSP, disassembly line balancing, and result evaluation.
Xia et al. propose a reinforcement learning framework with LLMs as an evaluator to explore optimized DSP results \cite{xia2024large}. A similar study by Ji et al. utilizes LLMs for sequence generation and reinforcement learning for workstation allocation to solve the disassembly line balancing problem \cite{ji2024reinforcement}. Another study by Xia et al. explores the capability of using LLMs' reasoning abilities to evaluate DSP results. They use a fine-tuned LLM with domain knowledge to ensure the robustness and reliability of the planning result \cite{xia2025leveraging}.
Subsequent studies further extend this direction to sequence reliability analysis \cite{hu2024human} and more efficient disassembly line balancing in complex settings \cite{guo2024large, guo2025llm}. These studies strengthen the planning results through LLM-based reasoning, but they still largely remain at the level of sequence generation and evaluation. Several recent studies have moved beyond planning by integrating perception, symbolic reasoning, and robotic control into the disassembly execution phase. Qi et al. \cite{qi2026llm} develop a hierarchical framework that utilizes LLMs and vision–language models (VLMs) for symbolic planning and state estimation, enabling robust, interpretable, and generalizable execution of long-horizon manipulation tasks in robotic disassembly. Liu et al. utilize the SAM and CLIP models for object detection and classification \cite{liu2024pretrained}. The detection results align with cloud-fog automation for task planning and execution in EV battery disassembly. Other studies have explored language-guided action planning and the use of prior execution experience to improve robotic disassembly \cite{peng2024revolutionizing, kang2025task, chang2025experience, moroncelli2025vision}. In HRC disassembly, LLM-based methods enable real-time coordination, interpretation, and safety-related reasoning that previously relied primarily on human intervention. The multi-modal information processing and knowledge retrieval capabilities of LLMs can support task allocation, understanding of human intentions, and adaptive planning during the HRC process.
Yu et al. propose a multi-modal LLM that integrates with a knowledge graph method for real-time task allocation in human-robot collaborative disassembly \cite{yu2025rescheduling}. Xiao et al. propose an LLM-guided graph neural network for real-time human intention prediction \cite{xiao2025large}. Tong et al. \cite{tong2026gnn} present a hybrid cognitive digital twin architecture that couples GNN-based interaction modeling with LLM-driven semantic inference, achieving self-adaptive task allocation and planning for multi-human and multi-robot collaborative disassembly.
Some studies have also demonstrated the effectiveness of using LLMs in real-time instruction generation and safety evaluation when considering HRC in disassembly \cite{kheirabadi2025llm, alenjareghi2025llm, alenjareghi2026proactive}. These studies suggest that LLM-related methods can extend disassembly practice from predefined and offline settings to real-time adaptive workflows that integrate planning, execution, and real-time coordination. However, robust automated disassembly under real-world EoL uncertainty remains in its early stages.

\subsection{LLMs in Recovery, Repair, and Reconditioning}

Following disassembly, components enter recovery, repair, and reconditioning, with subsequent treatments determined by their residual condition and the requirements of downstream remanufacturing operations. Cleaning is typically the first operation, since residual contamination may compromise subsequent assessment and downstream treatment decisions. Selecting an appropriate cleaning method is challenging because its effectiveness must be balanced against material compatibility and environmental constraints \cite{liu2013study, fadeyi2017integration}. Components are then repaired or replaced based on their functionality and required performance \cite{gharfalkar2016clarifying, wang2020optimization}. In some cases, obsolete parts may be substituted with upgraded alternatives, allowing the remanufactured product to achieve comparable or improved performance \cite{xing2007evaluation, copani2020remanufacturing, wu2022data}. Components that satisfy the relevant requirements are subsequently transferred to inventory before reassembly.

In practice, the execution of these operations rarely follows a fully standardized routine. These operations often require troubleshooting, task planning, and real-time adjustment under different component conditions. These rely not only on standard operating procedures (SOPs), but also on product conditions, operation instructions, and historical experience from human workers \cite{kanishka2023systematic}. In non-OEM remanufacturing, however, such knowledge is often incomplete or unavailable at the point of operation. Consequently, these processes remain heavily dependent on skilled workers. This dependence is labor-intensive and costly and may also introduce inconsistency in output quality. It may also introduce inconsistency in output quality \cite{jiang2019data}. Additionally, relying solely on humans exacerbates the mismatch between labor supply and the growing demand for remanufacturing \cite{zhu2025knowledge}.

This stage exposes a practical entry point for LLM-related methods. Recovery, repair, and reconditioning require workers to read and understand SOPs, repair records, and other process adjustment knowledge while making case-specific decisions during execution. Less-experienced workers often rely on guidance from senior workers to acquire task knowledge and resolve unfamiliar problems, creating an additional labor burden and inefficiency on experienced personnel. LLM-based systems could help ease this burden by providing timely knowledge support and stepwise guidance during operation. Recent work has started to move in this direction. Lu et al., for instance, explored VLMs for laptop reconditioning under condition uncertainty \cite{lu2025assessing}. The academic research in this area remains limited, partly because the operational data and domain knowledge required for LLM-related applications are difficult to acquire and structure under privacy and operational constraints. It is also difficult to rigorously demonstrate the value of applications in academic settings. In industry, however, companies can build internal applications around their own data and workflows, with strong potential to reduce reliance on experienced workers and lower the labor burden of knowledge-intensive operations.

\subsection{LLMs in (Re)Assembly and Test}

The distinctive challenge of reassembly in remanufacturing lies in confirming whether the components from prior recovery, repair, and reconditioning processes are still suitable for assembly. Components drawn from inventory may differ in condition, compatibility, and readiness for reassembly, especially when prior processes and reassembly take place in different facilities \cite{wlazlak2025development}. Reassembly, therefore, requires repeated confirmation of component suitability during operation, and component selection and reassembly sequence may need to be revised accordingly. In this case, it requires in-process decision-making and dynamic rescheduling, relying on human expertise and domain knowledge, both of which are labor-intensive and time-consuming. Testing is not a fixed final step either, as its procedures may need to change with component condition and reassembly outcomes.

Direct research on LLM-based reassembly in remanufacturing remains limited. Most related evidence comes from assembly tasks in manufacturing \cite{konstantinou2024leveraging, pang2025towards}. Existing studies suggest that LLM-related methods can support human–robot collaboration, task decomposition, and execution in complex assembly settings. Ma et al. \cite{ma2026probing} present an AR-assisted HRC assembly system integrated with multi-modal mutual cognition and LLM reasoning capability, enabling superior assembly performance in complex operational environments. Wang et al. \cite{wang2026llm} develop an LLM-driven HRC agent for complex aerospace wire-assembly tasks, demonstrating effective task decomposition and execution via the reasoning capability of LLMs. The advances in LLM-based assembly may inform reassembly, but the direct transfer from assembly to reassembly is constrained due to the uncertainty in the condition and suitability of components.

At this stage, LLM-related methods may retrieve and summarize records from prior remanufacturing processes to support reassembly and test decisions. This information can first guide operators to confirm the critical conditions of components, so that key uncertainties can be addressed early. Based on these checks, the system can interact with operators to support real-time decision-making and reassembly planning as needed. During reassembly, the system may continue to guide operators while recording observations. These observations can then be carried forward to adjust test planning, aligning the test procedure with reassembly outcomes and component condition. The LLM-related system can record test results, summarize the overall remanufacturing processes, and support the final assessment of the remanufactured product.

\subsection{LLMs in Design for Remanufacturing and Process Planning}

Design for remanufacturing and process planning are closely related in remanufacturing, but they operate at different levels. Both aim to improve the remanufacturing process based on feedback generated throughout the process. However, design for remanufacturing brings EoL information into the product design stage to reduce material use, cost, and remanufacturing complexity \cite{boorsma2022strategic, behtash2024reman}. Process planning in remanufacturing focuses on process selection, sequence planning, task planning, and operation scheduling.

In practice, design for remanufacturing depends on feedback from multiple perspectives, including customer feedback, process experience, financial analysis, and environmental regulations \cite{boorsma2021incorporating}. Aligning all information and achieving the design for remanufacturing goal remain challenging. Resolving these challenges requires integrating heterogeneous lifecycle information and feedback across multiple stakeholders, and existing design tools still offer limited support for this task. Current LLM-related studies have not focused directly on remanufacturing-oriented design, but they suggest useful directions for design for remanufacturing.
A recent study by Kumar et al. explored the potential of integrating LLMs in the CAD design stage with the FreeCAD module. LLMs can simplify traditional CAD workflows by allowing users to interact in natural language \cite{kumar2025generative}. Ashkbous et al. \cite{ashkbous2025leveraging} present a framework for integrating LLMs with product lifecycle data into a unified decision-support system for sustainable product development. Xiong et al. \cite{xiong2025dr} present a DR-RAG framework that incorporates domain knowledge graphs, rule-based reasoning, and digital-twin feedback with LLMs to enhance product design. Zhang et al. \cite{zhang2025multilingual} present a multilingual RAG framework with knowledge graphs for LLM-based reasoning and product design recommendation. 
Together, these studies show that LLMs can support knowledge integration and design reasoning, providing a useful basis for future design for remanufacturing-specific applications.

Process planning in remanufacturing still relies heavily on human expertise, product knowledge, troubleshooting ability, process capability, and facility capacity \cite{yazdanparast2025proposing, vahedi2025batch}. Prior studies have developed optimization-based methods for scheduling and related task planning.
Zhou et al. \cite{zhou2025multifactory} presented an optimization algorithm for designing work schedules across multiple remanufacturing facilities to improve efficiency and resource utilization. Yazdanparast et al. \cite{yazdanparast2025proposing} introduced a multi-agent reinforcement learning-based model to optimize the real-time phone remanufacturing scheduling problem. This model handles the arrival and processing times of different EoL parts. Liu et al. \cite{liu2026batch} presented a bi-directional hybrid multi-objective optimization model for the batched remanufacturing scheduling problem under the HRC setting. Similarly, Wang et al. \cite{wang2023hybrid} proposed a hybrid genetic algorithm incorporating variable neighborhood search to address the remanufacturing scheduling problem related to machine status for energy consumption optimization.
While these studies have demonstrated their effectiveness, they rely on structured inputs and well-defined assumptions. In practice, however, remanufacturing process planning depends on accumulated experience and operational knowledge drawn from large numbers of historical cases, which are difficult to formalize in the same way. This leaves a significant gap between existing computational models and human-level planning \cite{fu2024integrated, wang2021modeling} and motivates planning approaches that can directly leverage heterogeneous historical records.

Recent LLM-related work begins to address this gap by making heterogeneous historical records usable in process planning.
Zhang et al. established a remanufacturing process database that contains failure information and production history \cite{zhang2025llm}. They used LLMs to generate and optimize the remanufacturing process by retrieving the information from the established database. After that, the augment tool is used by the LLMs to calculate relative parameters that support the decision-making process. LLMs are also being used to interpret the optimization results for the remanufacturing process and to support real-time planning.
Current work mainly positions LLMs as planning assistants that use historical records, interpret outputs, and support real-time planning, rather than as fully autonomous planners.

\subsection{LLMs in Cross-stage and System-level Applications}

This section covers studies whose main contribution is not confined to a single remanufacturing stage. They are reviewed separately because they help advance remanufacturing toward higher levels of automation across multiple stages and, in some cases, at the system level. Their importance lies in reducing the human expertise that is still required for decision-making, planning, adaptation, and execution in variable remanufacturing settings.

In practice, remanufacturing still finds it difficult to move toward higher levels of automation, because many existing methods are developed for specific tasks and under defined settings. They often lack the generalization capability to handle uncertainty in EoL products \cite{de2025data, atuhurra2024leveraging}. Human expertise, therefore, remains necessary to evaluate the situation, perform decision-making, revise the plan, and adjust the operational process. This is particularly important when considering robotics in remanufacturing. The on-site operator may still be unable to translate a new requirement into the corresponding adjustment to the robot's behavior, leaving the automation dependent on additional expert intervention. The same difficulty is also seen beyond robotics, where changes in conditions still need to be understood and handled by experienced workers.

Existing studies already suggest how LLM-related methods may support this transition \cite{yu2026transformation}. Oyekan et al. \cite{oyekan2025applying} provide decision-making guidance on the use of LLMs, ontologies, and knowledge graphs in industrial automation, highlighting how LLM-based approaches may support HRC under more complex conditions. Additionally, LLM-based systems have also been studied for real-time detection, communication, decision-making, and dynamic planning \cite{tulbure2024study, jin2024reasoning}. Karli et al. \cite{karli2024alchemist} show that LLMs can support end-user robot application development through natural-language interaction, reducing the expertise required to translate operator intent into executable robotic behavior.
Zhou et al. \cite{zhou2026towards} present a modular VLM framework for zero-shot industrial robot tool manipulation, enabling structured reasoning and robust task execution without requiring task-specific tuning. Dalal et al. propose a Plan-Seq-Learn model that uses LLMs to guide reinforcement learning for solving robotic control tasks online \cite{dalal2024plan}. Liu et al. \cite{liu2026you} present an AR-assisted human–LLM collaborative motion-planning approach that leverages human-provided spatial information and LLM reasoning to achieve zero-shot motion planning for industrial mobile robots. While these studies suggest that LLM-related methods can reduce dependence on human expertise across remanufacturing operations, current work remains largely focused on discrete tasks rather than complete operational execution.
An extension of that is represented by vision-language-action (VLA) models, which directly map visual input to robotic actions as an end-to-end approach \cite{sapkota2025vision}. These models have shown strong generalization across manipulation and motion-related tasks \cite{kim2024openvla, lu2024research, wen2025tinyvla, qu2025spatialvla, o2024open, gao2025vla, abugurain2024integrating, driess2023palm}, suggesting a path toward reducing the manual effort still required to translate revised requirements into action. However, their usage in remanufacturing remains largely unexplored.

\begin{figure*}[htp!]
	\centering
	\includegraphics[width=0.95\textwidth]{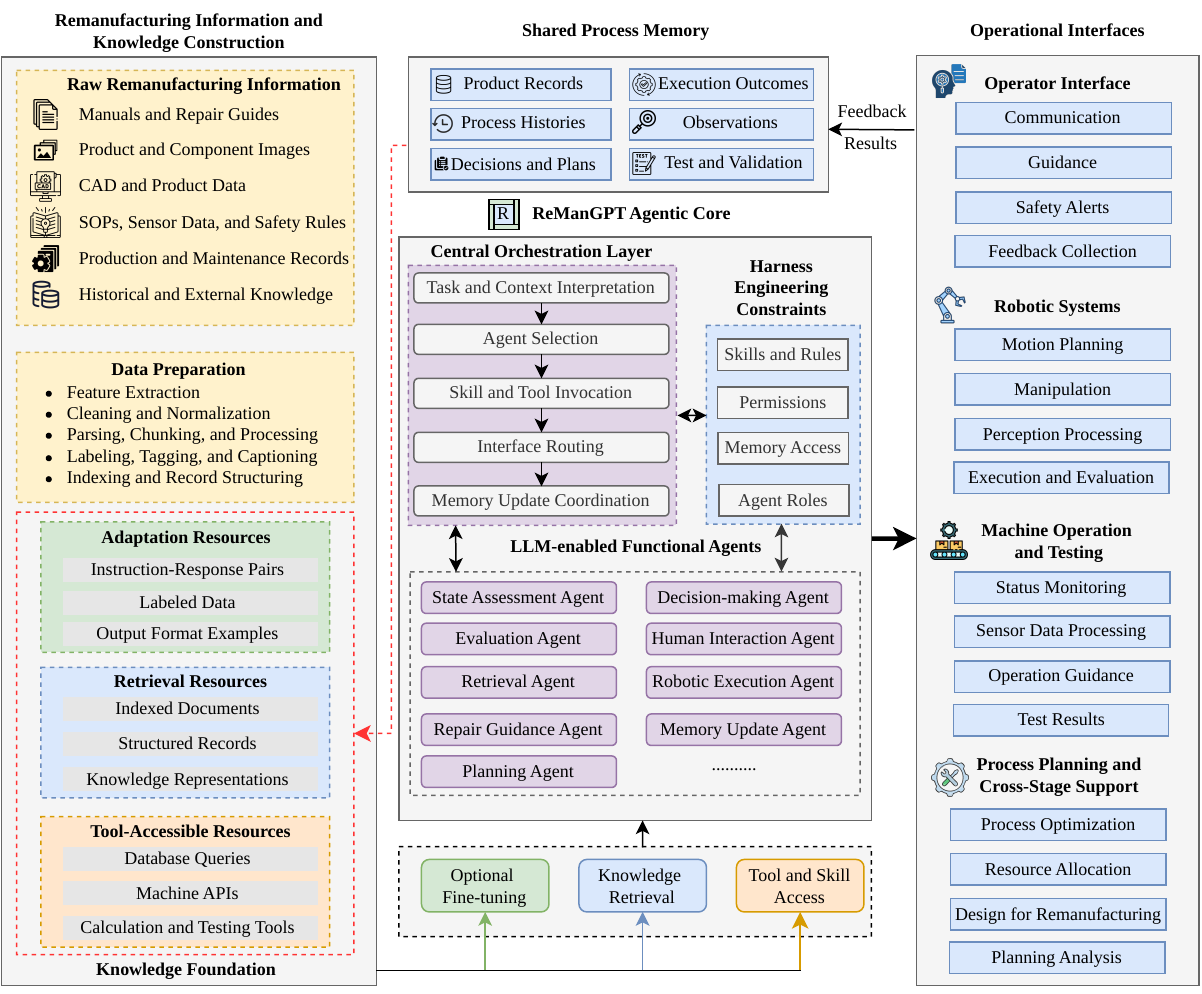}
        \vspace{3pt}
        \caption{Conceptual architecture and information flow of ReManGPT for remanufacturing. Heterogeneous remanufacturing information is processed into adaptation, retrieval, and tool-accessible resources. A central orchestration layer coordinates LLM-enabled functional agents under harness-engineering constraints and routes validated outputs to operational interfaces. Execution results, deviations, feedback, and validation outcomes are recorded in shared process memory, forming traceable product–component histories for subsequent decisions and future adaptation.}
	\label{fig:remangpt}
\end{figure*}

\section{ReManGPT: An Agentic Framework to Enhance Remanufacturing}

The review in the previous section shows that current LLM-related studies in remanufacturing remain largely task-specific. While these studies demonstrate the potential of LLMs in individual operations, they do not yet provide a coherent organization for linking knowledge, decisions, plans, and execution records across the remanufacturing workflow. To address this gap, ReManGPT is proposed as an agentic conceptual framework for LLM-enabled remanufacturing. It is defined at the framework level rather than as a single model, a standalone chatbot, or a completed industrial implementation. Its central orchestration layer manages module selection and information flow among LLM-enabled agents, shared process context, and operational interfaces, while the harness engineering structures and constrains agent behavior according to task requirements, safety rules, evidence grounding, and output formats.

This section presents the rationale, architecture, information flow, and operational workflow of ReManGPT. Three representative case studies then illustrate selected module-level instantiations in disassembly planning, human-centered repair assistance, and robotic execution, connecting the framework-level design to practical remanufacturing tasks.

\subsection{ReManGPT Framework}

\subsubsection{Framework Rationale}
ReManGPT is motivated by the fact that remanufacturing is executed through stages but governed by interdependent decisions. The inspection and sorting process establishes the initial basis for process routing by evaluating product condition, residual value, safety risk, and uncertainty. This assessment constrains the feasible disassembly strategy and determines the information needed for subsequent operations. Disassembly follows the strategy defined by inspection and determines the subsequent operation for each component. These component-level decisions further influence the remanufacturing target, reassembly plan, and testing requirements. Because each stage generates information that constrains later stages, remanufacturing cannot be adequately supported by isolated prediction, planning, or execution models. A stage-specific model may support a local operation, but its output remains limited if it is not connected with upstream evidence and downstream requirements. Remanufacturing, therefore, requires a coordination mechanism that can preserve process context, update decisions as product information becomes available, and transfer validated records across stages.

\begin{figure*}[!t]
    \centering
             \includegraphics[width=0.95\textwidth]{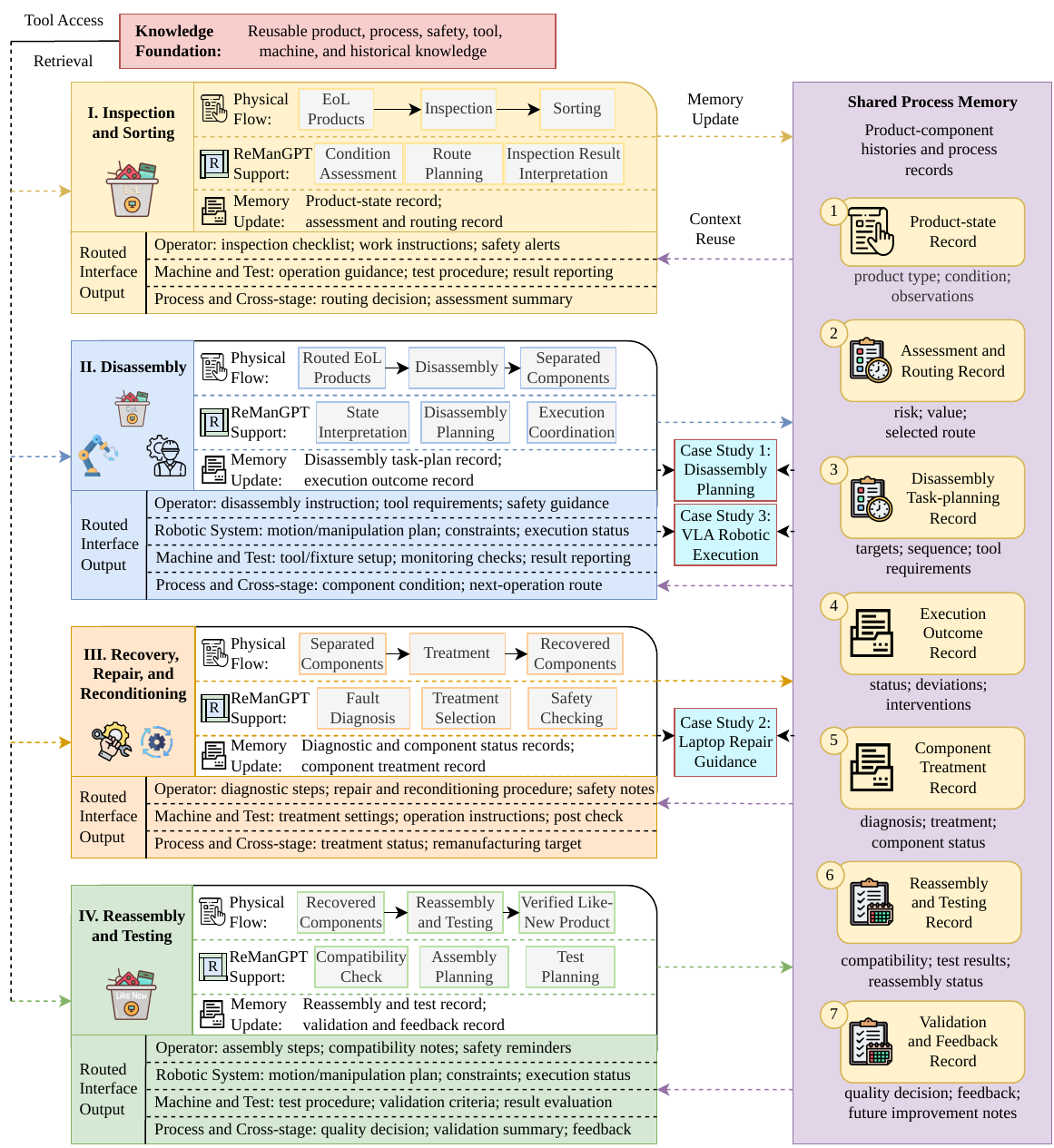}
            \vspace{3pt}
            \caption{ReManGPT-enabled remanufacturing workflow with knowledge retrieval, operational interface routing, and shared process memory. It shows how reusable knowledge is accessed through retrieval and tool access, how ReManGPT supports each remanufacturing stage, and how outputs are routed to different interfaces. Stage outcomes are written to shared process memory as product–component records and reused as context by downstream stages. The case labels mark the positions of the three representative module-level case studies.}
    \label{fig:operalevel_1}
\end{figure*}

ReManGPT is introduced to provide this coordination at the framework level. It organizes stage-specific LLM-enabled agents around shared process context and a central orchestration layer, enabling assessments, plans, execution outcomes, and feedback to be carried across the remanufacturing workflow. The conceptual architecture of ReManGPT is illustrated in Fig.~\ref{fig:remangpt}. Building on this architecture, Fig.~\ref{fig:operalevel_1} presents the operational workflow, showing how ReManGPT can support decision-making and process execution through different remanufacturing operational stages.

\subsubsection{System Architecture}
ReManGPT is structured around a central orchestration layer that connects four architectural elements: the remanufacturing knowledge foundation, shared process memory, LLM-enabled functional agents, and operational interfaces. The knowledge foundation represents the reusable domain resources required by the framework, including product, process, safety, and historical knowledge. In contrast, shared process memory represents the case-specific context associated with a particular EoL product, component, or task. This separation allows ReManGPT to rely on established remanufacturing knowledge while maintaining the evolving context of individual product instances.

The orchestration layer forms the organizing core of the framework. It does not replace specialized models or tools but provides a common structure for selecting, composing, and connecting them to the current remanufacturing context. Within this structure, functional agents are defined by their operational roles, such as assessment, retrieval, planning, repair guidance, human interaction, robotic execution, verification, and memory management. These functional agents are coordinated modules because their outputs are interpreted through shared context and directed toward specific operational targets, rather than toward isolated, task-specific solutions.

Harness engineering provides the constraints that make this agentic architecture usable in remanufacturing settings. It defines the role of each agent, the scope of its accessible context, the skills and tools it may use, the expected output format, and the verification requirements associated with its task. In this sense, the orchestration layer determines the structural coordination among agents, knowledge, memory, and interfaces, while harness engineering ensures that agent behavior remains aligned with task requirements, safety rules, evidence grounding, and downstream usability. The operational interfaces define the execution boundary of ReManGPT. They translate validated outputs from the orchestration layer into execution-level representations that operators, robotic systems, machines, testing equipment, or process-planning systems can follow, execute, or check in the remanufacturing environment. After execution, observed outcomes, deviations, and user feedback are returned through these interfaces to the orchestration layer and linked to the relevant product or component record in shared process memory.

\subsubsection{Information Flow and Adaptation}
The information flow in ReManGPT begins with the construction of the knowledge foundation. Remanufacturing information is originally heterogeneous and often unstructured, including manuals, repair guides, product images, CAD files, production records, inspection reports, historical cases, and external knowledge. Before it can be reliably used by LLM-enabled agents, this information must be curated and processed into high-quality, agent-accessible resources. Text documents may be cleaned, parsed, chunked, indexed, and annotated. Images and CAD-related information may be processed and linked with product or component descriptions. Operational information may also be normalized into product, component, task, or outcome records. Through this preparation, raw information is transformed into structured, retrievable knowledge resources that support later reasoning, planning, and execution.

These prepared resources support two complementary forms of adaptation. Curated instruction-response data can be used for fine-tuning, allowing the model to better follow remanufacturing terminology, procedural patterns, and task-specific output formats. In parallel, indexed documents, structured records, and knowledge representations can be formed as databases to support knowledge retrieval. This allows agents to access updated evidence, case-specific constraints, and prior operational records during inference. Fine-tuning adapts model behavior, while retrieval and tool access provide task-relevant information at runtime. Additionally, with the development of the ReManGPT framework, the information flow no longer relies solely on manually prepared structured data. Functional agents equipped with appropriate skills can access semi-structured or unstructured resources through controlled retrieval, parsing, vision-language interpretation, CAD queries, database search, or other tool-use mechanisms. In this setting, an agent may not only retrieve existing structured knowledge but also convert newly encountered raw information into structured records for storage and reuse. Harness engineering remains important in this process because it constrains what information an agent can access, how it can process that information, and what form the generated record must take.

The outputs generated by agents also become part of the information flow. An assessment result, disassembly plan, repair instruction, verification result, or execution command is associated with the current product, component, or task context. When an output is applied through an operational interface, the execution result either confirms the current process state or reveals a deviation that must be retained. The returned evidence is recorded in shared process memory as part of the product–component history. As the same product moves through subsequent operations, this history allows downstream agents to interpret conditions, revise plans, and justify subsequent decisions based on prior evidence. Validated records may also be retrieved for similar future cases or curated for later adaptation. This information flow and adaptation design make the ReManGPT framework traceable for each product instance and adaptive across future cases, because each validated operation contributes to the product–component history that guides subsequent agents.

\begin{table*}[htp]
\small
\centering
\caption{Representative case studies as selected ReManGPT module instantiations.}
\label{tab:remangpt_case_mapping}
\setlength{\tabcolsep}{4pt}
\renewcommand{\arraystretch}{1.12}
\begin{tabularx}{\textwidth}{@{}
>{\raggedright\arraybackslash}p{2.55cm}
>{\raggedright\arraybackslash}p{3.05cm}
>{\raggedright\arraybackslash}p{4.05cm}
>{\raggedright\arraybackslash}X
@{}}
\toprule
\textbf{Case study}
& \textbf{ReManGPT module}
& \textbf{Main information used}
& \textbf{Generated record and framework connection} \\
\midrule

Disassembly sequence planning
& Planning module before physical disassembly
& Product images, user query, component-value information, tool knowledge, and disassembly constraints
& Component targets, blocking relations, disassembly sequence, and tool requirements; stored as a disassembly task-plan record in shared process memory and routed to human or robotic execution. \\

\midrule

Laptop repair and replacement assistant
& Repair-guidance module during recovery, repair, and reconditioning
& Failure description, user-provided diagnostic findings, repair-guide knowledge, tool requirements, and safety constraints
& Diagnostic procedure, repair instruction, and safety notes; diagnostic and component-treatment records are generated and stored in shared process memory for downstream verification. \\

\midrule

VLA-based robotic disassembly
& Robotic-execution module for selected component-removal tasks
& Runtime observations, robot state, language instruction, and paired demonstration data for adaptation
& Predicted robot actions and execution outcomes; stored as an execution record for later assessment, replanning, or human intervention. \\

\bottomrule
\end{tabularx}
\end{table*}

\subsubsection{Operational Workflow}
The operational workflow describes how the preceding architecture and information flow are applied in remanufacturing practice. In operation, the orchestration layer follows a defined coordination procedure. It first interprets the task request together with the current product or component context stored in shared process memory. It then selects the relevant functional agent and determines the knowledge, tools, and interface needed for the task. The selected agent retrieves evidence or invokes tools under the constraints defined by harness engineering, and then generates a structured output such as an assessment, plan, instruction, or execution command. This output is checked against safety rules, task requirements, and available evidence before it is routed to the operator, robot, machine, testing system, or process-planning interface. After execution, the result, deviation, feedback, or validation outcome is written back to shared process memory. This procedure allows ReManGPT to coordinate modules without treating any single LLM output as an unchecked final decision.

As shown in Fig.~\ref{fig:operalevel_1}, an EoL product first enters inspection and sorting, where available observations, records, and domain knowledge are organized into an initial assessment and process route. This route provides the starting point for downstream operations, but it remains subject to revision as new product information becomes available. During disassembly, hidden structures and component conditions are gradually exposed. These observations can confirm the initial assessment or reveal deviations that require updating the process route. The orchestration layer can then revise task plans, request additional inspection, or direct the next operation to the appropriate operational interface. The generated outputs may take the form of operation-level guidance, structured task records, or language instructions, depending on whether the next step is performed by an operator, a robotic system, or another process module. In this way, disassembly is not treated as a fixed execution of a prior plan, but as a stage in which product and component context are progressively refined and translated into executable operational steps. After disassembly, dismantled components carry their accumulated condition history into recovery, repair, and reconditioning. ReManGPT can use this history, along with process constraints and available knowledge, to support treatment selection, repair guidance, and documentation. The outcomes of these treatments then become part of the same product–component history and inform reassembly and testing. Reassembly decisions can therefore reflect not only the original product assessment, but also the treatment route, compatibility, and verification status of individual components.

\begin{figure*}[htp!]
	\centering
	\includegraphics[width=0.9\textwidth]{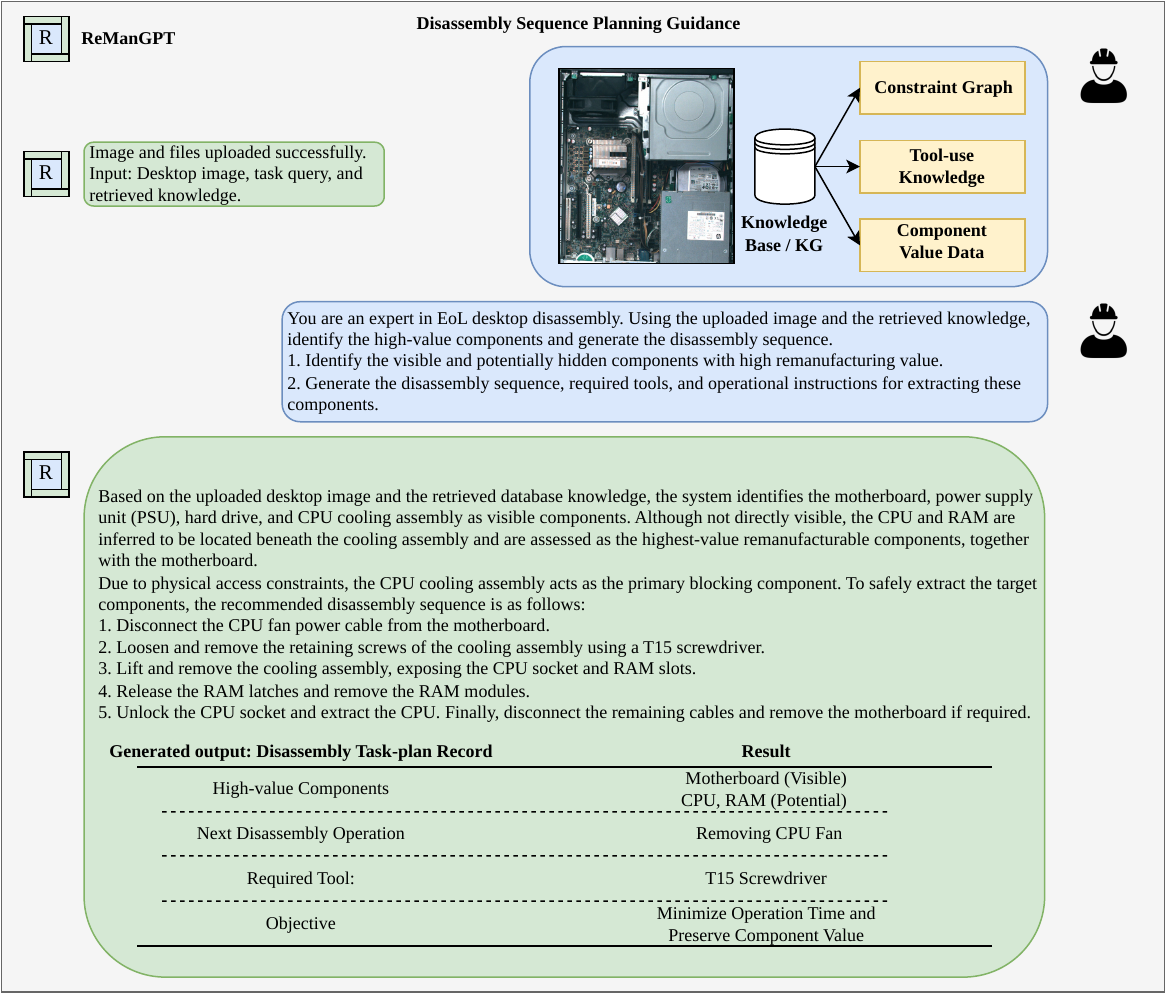}
        \vspace{0pt}
        \caption {Case study of the ReManGPT disassembly-planning module using a multi-modal LLM and KG-based knowledge retrieval. The generated output is a disassembly task-plan record containing target components, blocking relations, sequence, and tool requirements.
        }
	\label{fig:DSP_CASE}
\end{figure*}

Across these stages, the role of ReManGPT is to coordinate information, module outputs, and operational interfaces rather than to replace all physical operations. Plans, instructions, execution outcomes, and validation results are carried forward as part of the shared process context, allowing later decisions to be made with reference to prior evidence. The workflow, therefore, provides a closed-loop operational abstraction for module coordination under changing EoL product conditions. The same operational records can also support higher-level remanufacturing functions. Accumulated evidence on disassembly difficulty, treatment feasibility, resource demand, and verification outcomes can inform process planning across products and batches. It can also reveal recurring product features that hinder inspection, disassembly, repair, or final verification, providing feedback for design for remanufacturing. 

Taken together, this operational workflow describes how ReManGPT carries product and component context from assessment through execution to verification. Its role is to keep shared evidence, updated plans, operation-level guidance, and validation results connected as EoL product conditions change. Rather than prescribing a fully autonomous implementation, the workflow specifies the coordination logic through which selected agents, interfaces, and process records can be developed, validated, and integrated under human and production oversight.

\subsubsection{Case Study Positioning}
The case studies are positioned as selected module-level instantiations of the ReManGPT architecture. They cover three points at which framework outputs must be used in different forms: planning before execution, operator guidance during component treatment, and robotic execution. Each case is therefore described by the information it consumes, the record or output it generates, and its connection to the shared process context.

Table~\ref{tab:remangpt_case_mapping} summarizes the module role, information use, generated memory record, and framework connection for each case. The case outputs are interpreted through the shared process memory and operational interfaces, rather than as stand-alone results. In the workflow shown in Fig.~\ref{fig:operalevel_1}, the disassembly-planning case creates a task-planning record that can be routed to an operator interface or a robotic execution interface. The VLA-based robotic disassembly case illustrates how an execution module uses task-level instructions, visual observations, and robot-state data to predict robot actions and return execution outcomes. The repair-assistance case creates diagnostic and treatment records that become part of the component history and support later verification, reassembly, or testing. These links place the three cases at different points in the same ReManGPT workflow. Although the demonstrations are evaluated separately, the framework defines the memory records and operational interfaces through which they can be connected.

\subsection{Case Studies}
Building on the framework-level positioning above, this section turns to the task-level demonstrations of the selected modules. The purpose is to examine how each module operates for a concrete remanufacturing request, what information it uses, what output it generates, and what limitations arise in practice. These details provide the basis for assessing the feasibility and current maturity of the corresponding ReManGPT modules.

\begin{figure*}[htp!]
	\centering
	\includegraphics[width=0.9\textwidth]{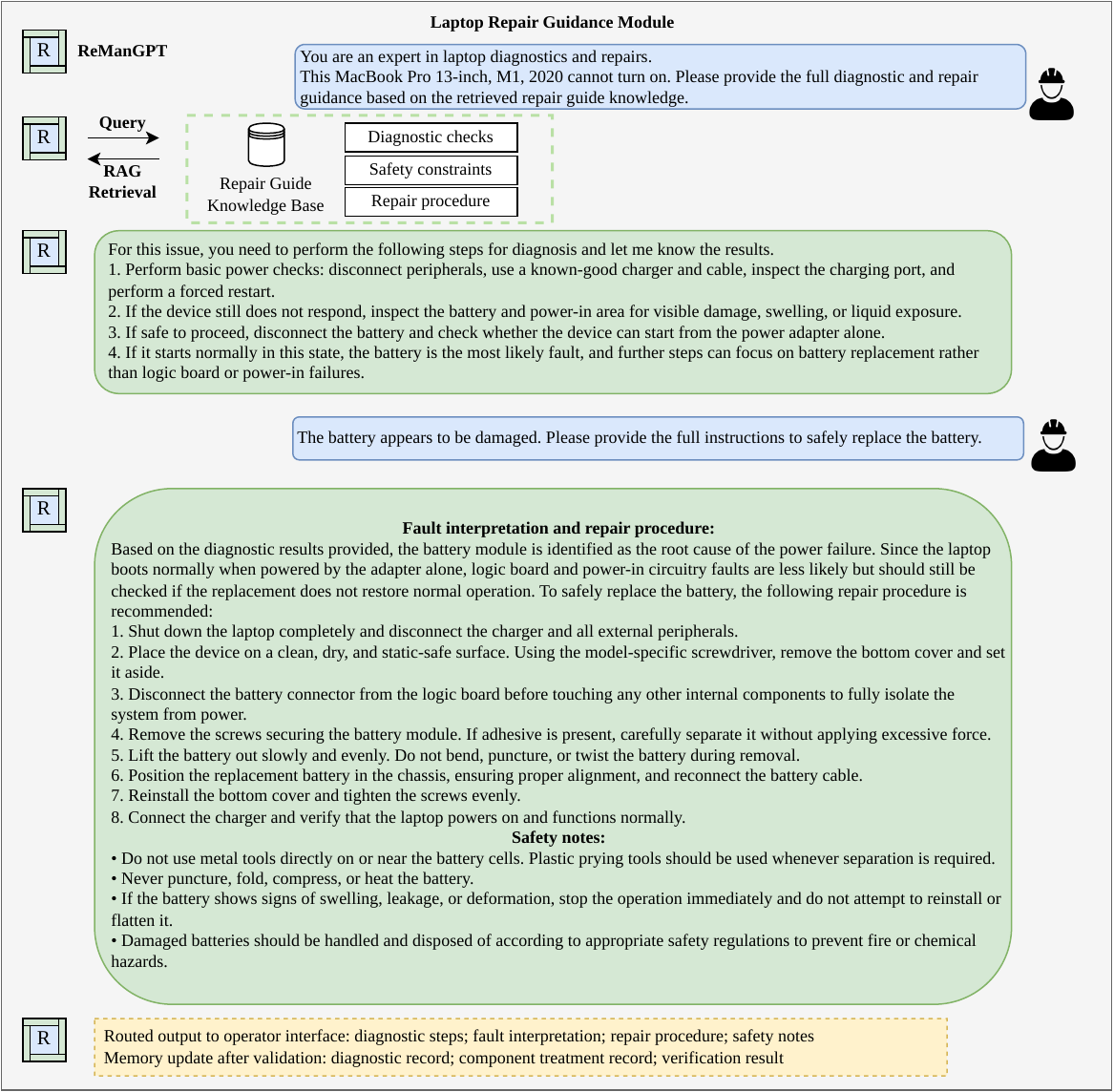}
        \vspace{0pt}
        \caption{Case study of the ReManGPT repair-guidance module using LLM and RAG-based repair-guide retrieval. The module routes diagnostic guidance, repair procedures, and safety notes to the operator interface and, after validation, updates shared process memory with diagnostic and component treatment records.
        }
	\label{fig:RA_Case}
\end{figure*}

\subsubsection{Disassembly Sequence Planning}
This case demonstrates the disassembly-planning module of ReManGPT using an EoL desktop disassembly task. The module operates before physical execution and converts product observations and retrieved domain knowledge into component identification, target selection, and sequence-level planning. This task is representative of selective disassembly because high-value components may not be directly visible and can be blocked by other parts under uncertain product conditions. The module is adapted using a customized dataset collected from 10 desktops, including more than 2800 image–query–response pairs covering component identification and disassembly procedures.

The test case involves a desktop with a similar internal structure, but it is not included in the fine-tuning dataset. During inference, the module receives an image of the open desktop and a task-specific query to identify high-value components and generate a disassembly plan. It also retrieves domain knowledge from the database, including component value, tool usage, and disassembly constraints. In the ReManGPT framework, these inputs correspond to the product observation, task request, and knowledge foundation available to the planning module. The generated output includes identified visible components, potentially hidden components, high-value targets, and a corresponding disassembly sequence with operation and tool requirements, as illustrated in Fig.~\ref{fig:DSP_CASE}.

In this case, the module identifies visible components such as the motherboard, power supply, storage device, and CPU cooling system. It also infers that the CPU and RAM, which are treated as high-value targets, may be blocked by the cooling system and storage drive. Based on this assessment, the module generates a sequence-level plan that removes the blocking components before accessing the target components and specifies the required tools and operations. The resulting component targets, blocking relations, sequence plan, and tool requirements form a task-planning record in ReManGPT. This record can be retained in shared process memory and used to guide later human operation or provide task-level input for a robotic execution module. This case, therefore, illustrates how a ReManGPT planning module can transform limited product observations and domain knowledge into an operation-ready planning record. It also shows the current boundary of the module. Planning failures are observed when desktop layouts differ substantially from those represented in the adaptation data, indicating that the current module remains limited in generalizing to unfamiliar EoL configurations. In the broader ReManGPT workflow, such failures would need to be captured as planning uncertainty or feedback for additional observation, retrieval, or human review before execution.

\begin{figure*}[htp!]
	\centering
	\includegraphics[width=0.9\textwidth]{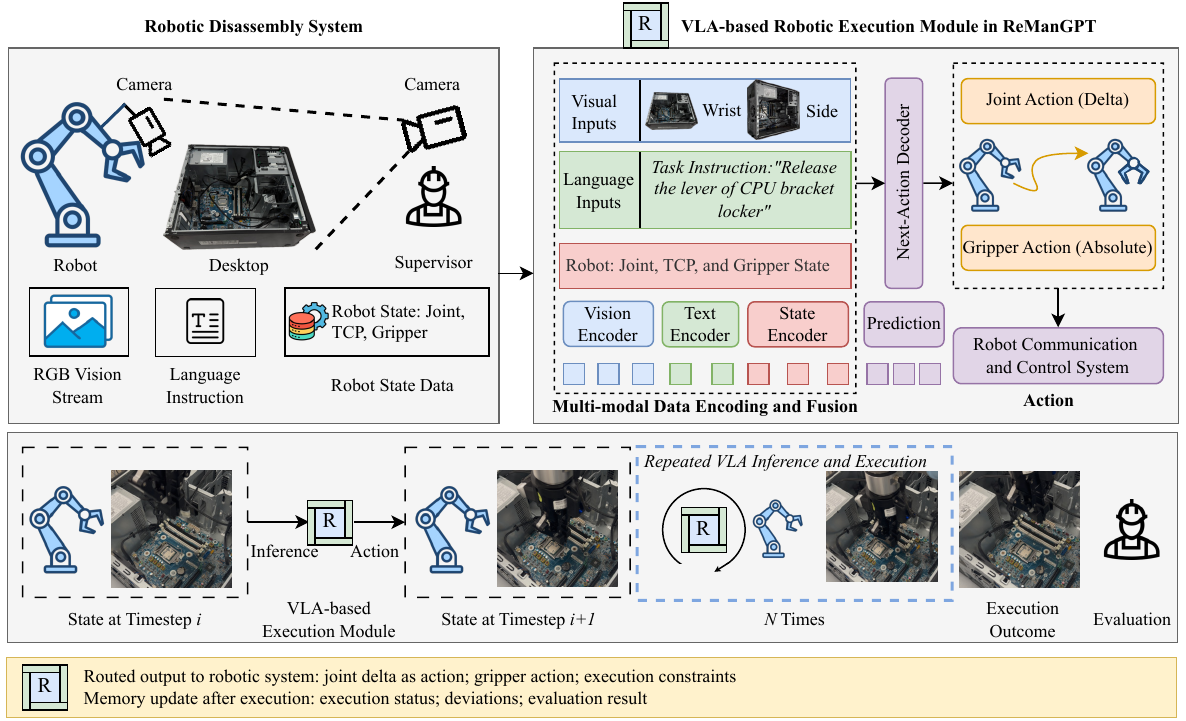}
        \vspace{0pt}
        \caption{Case study of the ReManGPT robotic-execution module for VLA-based EoL desktop disassembly. The module converts multiview visual inputs, task instructions, and robot-state data into closed-loop robot actions, while recording execution outcomes in shared process memory for later assessment and replanning.}
	\label{fig:VLA_Case}
\end{figure*}

\subsubsection{Laptop Repair and Replacement Assistant}
This case examines a repair guidance module of ReManGPT for laptop diagnosis and component replacement.  The module is positioned in recovery, repair, and reconditioning, where operators must interpret failure symptoms, identify likely causes, and select repair actions under tool and safety constraints. This task is knowledge-intensive because diagnostic logic and repair procedures are often distributed across repair guides, product-specific instructions, and practical experience. To adapt the module to this setting, a structured dataset of diagnostic and repair instructions was collected from iFixit repair guides for various MacBook laptop models. The dataset includes more than 340 text-based query–completion pairs covering symptom descriptions, diagnostic checks, repair steps, tool requirements, and safety constraints.

The demonstration uses an in-domain MacBook Pro power-on failure scenario within the repair knowledge scope represented by the curated dataset. During inference, the module receives a failure description and a diagnostic request from the user. It first generates a structured diagnostic procedure to narrow down the possible causes. After the user follows the diagnostic procedure and reports that the battery is the likely failure source, the module generates a battery replacement procedure with ordered steps, tool requirements, and safety precautions, as illustrated in Fig.~\ref{fig:RA_Case}. The generated output consists of diagnostic guidance, fault interpretation, and stepwise battery-replacement instructions, including tool requirements and safety precautions. It is intended to support operator execution and remains subject to physical verification during repair.

Within ReManGPT, this module converts failure descriptions, user-provided diagnostic findings, and repair-guide knowledge into structured repair guidance and traceable repair records. Once validated during operation, these records can be retained as part of the component history and used in later reassembly, testing, or quality verification. This case, therefore, illustrates how a ReManGPT repair-guidance module can reduce the knowledge burden on operators while preserving the diagnostic and repair evidence needed by downstream operations. The limitation of this module lies in its dependence on the breadth and quality of the underlying repair knowledge. When device models, failure modes, tool requirements, or safety conditions are not adequately represented, the generated guidance may require additional retrieval, expert review, or physical confirmation before use. This limitation is consistent with the role of the module as operator support rather than autonomous repair execution.

\subsubsection{VLA-based Robotic Disassembly System}
This case examines a robotic execution module of ReManGPT for selected EoL desktop disassembly tasks through vision-language-action (VLA) models, as illustrated in Fig.~\ref{fig:VLA_Case}. The module is positioned at the execution level of disassembly, where visual observations, language instructions, and robot states are converted into predicted robot actions. The study focuses on CPU and RAM removal, which represent component-removal tasks that require target localization, constrained motion, and contact-rich manipulation within a compact desktop structure.

During execution, the robotic system provides multi-view visual observations, the controller provides robot-state information, and a short language instruction specifies the task. The VLA models interpret these inputs, form an understanding of the current state, and predict the next robot action, including joint movements and gripper configurations. The process is closed-loop in the sense that each predicted action changes the scene and produces new observations for the next prediction. In the current prototype, the language instruction is provided by a human supervisor. Within the ReManGPT framework, this instruction corresponds to the type of task-level record that can be generated by an upstream planning module, such as the disassembly-planning module discussed earlier. The present case, therefore, examines the execution-side module rather than a fully integrated planning-to-execution pipeline.

During the data collection phase, we utilize the Gello \cite{wu2024gello} platform to collect robotic disassembly data remotely across ten different desktops. A total of 165 sets of data were collected for RAM disassembly, and 123 sets of data for the CPU removal process. The control frequency and camera recording frequency are set to 30 Hz for alignment purposes. The fine-tuning datasets are constructed by pairing aligned perceptions and state information with the corresponding robot actions. During the evaluation process, we select two desktop configurations that are shown in the collected datasets. The physical evaluation shows that the fine-tuned models can localize the target and approach it with the intended gripper configuration, but they do not complete the full pick-and-disassemble sequence. This result indicates that the module learns some aspects of pre-contact behavior, including target approach, gripper alignment, and configuration maintenance, but remains insufficient for reliable component removal after contact. The gap reflects the difficulty of robotic disassembly of EoL products, where components are closely arranged, visibility is limited, and manipulation often requires force-sensitive contact. Force and tactile sensing would be valuable for handling these conditions \cite{lei2026learning}, but they are not included in the selected VLA models. Historical state information is also not explicitly integrated, although it may be important for long-horizon disassembly tasks.

Within ReManGPT, this case identifies the execution bottleneck between task-level planning and reliable physical disassembly. The module predicts robot actions and produces execution outcomes, but an unsuccessful removal is also informative, as it becomes an execution record associated with the attempted task, product state, and observed failure mode. Such records can be returned to shared process memory and used to support later assessment, replanning, human intervention, or model adaptation. The value of this case is therefore not to claim full robotic disassembly, but to show how a robotic execution module can be evaluated within ReManGPT and how its limitations reveal the additional information required for execution, including tactile feedback, force sensing, improved visual coverage, and more detailed task instructions. We provide a broader discussion of these challenges and the potential research direction in the future direction section.

\section{Focused Applications}

In this section, we focus on three applications highlighted in the introduction. We summarized their uniqueness in the remanufacturing domain from different perspectives in Table~\ref{table:REapplication}. The conceptual framework ReManGPT, which we proposed earlier, offers a novel approach to current remanufacturing practices. Across all the applications, ReManGPT can initially reduce dependence on human expertise by retrieving relevant knowledge from design, usage data, and historical records, enabling human workers to communicate with ReManGPT to acquire interpretable instructions that address uncertainties. It can also serve as a system-level decision-support and orchestration layer that coordinates existing machine learning tools, analytical models, automation systems, human operators, and robotic platforms to provide real-time, context-aware decision support throughout the remanufacturing workflow. As the automation level increases, ReManGPT can translate selected decisions and plans into executable instructions or robotic commands, enabling robots to safely handle complex tasks while collaborating with humans or, in some cases, operating with reduced human involvement in the remanufacturing process.

\begin{table*}[htp]
\small
    \centering
    \caption{Three remanufacturing applications with their domain features, special objectives, and challenges.}
    \label{table:REapplication}
    \begin{tabular}{>{\raggedright\arraybackslash}m{1.3cm}
                    >{\raggedright\arraybackslash}m{3.8cm}
                    >{\raggedright\arraybackslash}m{5cm}
                    >{\raggedright\arraybackslash}m{5.5cm}}
        \toprule
        \textbf{Application} & \textbf{Domain features} & \textbf{Objectives} & \textbf{Challenges} \\
        \midrule

        \textbf{EV-LIBs}
        & {\raggedright
           \textbullet~High economic value and strict safety constraints. \newline
           \textbullet~Structural and state uncertainty from design diversity and aging conditions. \par}
        & {\raggedright
           \textbullet~Estimate SOH/SOC/RUL under incomplete lifecycle information. \newline
           \textbullet~Provide safety-aware and interpretable operational guidance. \newline
           \textbullet~Adaptive disassembly planning for different designs. \par}
        & {\raggedright
           \textbullet~Diagnostic uncertainty makes the process heavily reliant on expert judgment. \newline
           \textbullet~Disassembly requires human expertise to handle uncertainty from different joining methods. \newline
           \textbullet~Current automation cannot handle these uncertainties, so extensive manual work increases safety risks for operators.\par}
        \\
        \midrule
        \textbf{E-waste}
        & {\raggedright
           \textbullet~Low per-unit value but extremely high processing volume.\newline
           \textbullet~Uncertainty from large variation in product types, designs, conditions, and generations.\par}
        & {\raggedright
           \textbullet~Identify and select high-value components for disassembly. \newline
           \textbullet~Determine optimal disassembly level and sequence, then sort components correctly. \newline
           \textbullet~Maintain high throughput and efficiency to ensure profitability. \par}
        & {\raggedright
           \textbullet~Complex uncertainties make identification, planning, and sorting heavily dependent on human decision-making.  \newline
           \textbullet~Selective disassembly and component sorting rely on manual operation. \newline
           \textbullet~Current automation systems cannot handle the uncertainties and perform the disassembly tasks, making the whole process labor-intensive and unprofitable. \par}
        \\
        \midrule
        \textbf{Electric motors}
        & {\raggedright
           \textbullet~High retained value in core components that contain rare earth elements and often remain in good condition. \newline
           \textbullet~Large variation in motor sizes, types, usage scenarios, and EoL conditions, resulting in diverse remanufacturing methods. \newline
           \textbullet~Compact structural designs with tight tolerances and limited internal visibility before disassembly. \par}
        & {\raggedright
           \textbullet~Perform component assessment for various electric motors during and after the disassembly process to determine appropriate reconditioning routes. \newline
           \textbullet~Provide instructions for non-destructive and precise disassembly operations to preserve the value of core components. \newline
           \textbullet~Offer accurate reassembly guidance that is based on component condition and the selected reconditioning processes. \par}
        & {\raggedright
           \textbullet~Internal conditions and failure modes cannot be fully evaluated before disassembly, making component assessment dependent on human expertise. \newline
           \textbullet~Large variation in types and structures makes non-destructive and precise disassembly highly dependent on manual operations performed by experienced workers. \newline
           \textbullet~Automation cannot meet the precision and dexterity requirements for these fine-grained disassembly and reassembly tasks.\newline
           \textbullet~Reconditioning decisions rely on operator experience due to limited and unstructured data that may support the process. \par}
        \\
        \bottomrule
    \end{tabular}
\end{table*}

\subsection{Electric Vehicle Lithium-ion Batteries}
Energy storage systems, particularly lithium-ion batteries in electric vehicles (EV-LIBs), have attracted global attention due to their growing demand and early EoL challenges \cite{neri2024sustainable, li2018cost}. Global EV sales rose by 27\% in 2024, reaching 17.5 million units, and are expected to reach 40 million by 2030 \cite{IEA_Global_EV_Data_Explorer_2025}. This surge is expected to bring the total EV battery demand to 3 TWh by 2030 \cite{IEA_Global_EV_Outlook_2025_Batteries}. In particular, EV-LIBs are generally considered to have reached the EoL stage when their state of health (SOH) falls below 80\% \ of the original capacity. These EoL EV-LIBs retain a large portion of high-value functional components. Since the production of new batteries remains highly dependent on critical resources such as REEs, remanufacturing stands out as a high-potential recovery approach. With the increasing number of EVs, the potential recovery market for EV batteries is projected to reach \$18.1 billion by 2030 \cite{meng2022intelligent}.

Although EV-LIBs represent some of the most valuable targets in remanufacturing, their practical processing remains challenging because uncertainty permeates every stage of the workflow \cite{kamath2023system}. The inspection and sorting process requires an accurate diagnosis of the state of health (SOH), state of charge (SOC), and remaining useful life (RUL), which is challenging because the collected EoL EV-LIBs often lack lifecycle data and contain limited information, resulting in highly variable patterns. These facts make the inspection and sorting process inefficient, dependent on experts, and prone to misclassifications. In the disassembly stage, the diagnostic results from the inspection, combined with the variety of battery pack designs, internal conditions, and joining techniques, make the disassembly process highly uncertain. In practice, real-time reasoning and cognitive capabilities are required for dynamic and adaptive high-quality disassembly operations. The recovery and reassembly stages also rely heavily on expertise, requiring proper knowledge retrieval and process recording, both of which are currently time-consuming for human workers. These factors not only limit the scalability and efficiency but also expose human workers to a potentially unsafe working environment due to high voltage, hazardous materials, and risky operations \cite{neri2024sustainable, meng2022intelligent}.

Recent advances in machine learning and robotics for intelligent EV-LIBs remanufacturing have yielded promising capabilities, including automated target detection, component selection, dynamic disassembly planning under uncertain conditions, process optimization, disassembly automation \cite{huang2026robotic}, and structured forms of human–robot collaboration \cite{tan2025robotic}. Although these approaches have achieved impressive results within their respective task boundaries, they remain isolated, and no existing system integrates these capabilities into a unified and end-to-end workflow \cite{hertel2024towards}. Additionally, most current models lack interpretability, requiring technicians to manually validate outputs by cross-referencing prior data and documenting operational details themselves, which adds workload and constrains practical deployment. These limitations underscore the need for a more comprehensive and information-driven framework, which is capable of supporting scalable and intelligent EV-LIBs remanufacturing.

ReManGPT could provide a structured decision-support layer to address these challenges in EV-LIBs remanufacturing. By retrieving information from design profiles, regulatory standards, and historical records, it enhances diagnostic reliability, which is the most important factor in determining the value of EV-LIBs remanufacturing. Workers without specialized expertise could interact with such a system to obtain clear and interpretable instructions. Through natural language communication, ReManGPT can explain model outputs, adjust plans based on new observations, and document process details, thereby reducing the overall human workload. It could also serve as an orchestration layer to coordinate machine learning models, analytical tools, automation systems, operators, and robotic platforms, thereby handling uncertainty and achieving a higher level of automation in remanufacturing. ReManGPT could support safer robotic involvement by providing reasoning and planning support, which may reduce safety risks and improve overall efficiency throughout the EV-LIBs remanufacturing process.

\subsection{Electronic Waste}

Electronic waste (e-waste) has become one of the fastest-growing waste streams globally, driven by the rapid replacement of products and technological advancements. In 2022, global e-waste reached 62 million tons, with only 22.3\% properly recycled \cite{Globalewaste2024}. Unlike EV-LIBs and electric motors, which have a high remanufacturing value, e-waste stands out for its extraordinary volume. Low collection rates and poor waste management strategies result in significant losses of valuable components and materials, including REEs. Additionally, current recycling practices raise ongoing concerns about their environmental impact. Remanufacturing offers a more sustainable solution for e-waste recovery \cite{lee2024environmental}.

E-waste presents a different set of challenges for remanufacturing compared to the EV-LIBs, which have significantly high values. The rapid iteration of technologies and the growth in product variety lead to high uncertainty in internal structures and usage conditions. Unlike EV-LIBs, e-waste has no standardized evaluation metrics for assessing EoL condition, making real-time adaptive decision-making and planning important throughout the remanufacturing process \cite{lee2022task}. At the same time, the extraordinary volume of e-waste creates pressure from both environmental and operational perspectives. However, its relatively low per-unit remanufacturing value means that the whole process is only profitable when the overall throughput is high, and labor involvement is kept to a minimum. These factors drive the need for a low-cost, easily deployable, and adaptive automated system in decentralized settings. Such automated systems must handle diverse product categories and significantly improve efficiency in order to make e-waste remanufacturing economically sustainable \cite{liu2026raise}. Recent studies have shown that remanufacturing automation is partially feasible under controlled conditions \cite{liu2024hybrid}, but these methods remain isolated among different tasks and are hard to generalize in real-time settings.

ReManGPT offers a potential approach to support decision-making and information consistency in e-waste remanufacturing, thereby improving efficiency. While robotics development is still far from being mature in practice, the ReManGPT framework can support the decision-making and planning process in the current human-centric e-waste remanufacturing process. The current practice still relies on human expertise to identify targets, determine disassembly scopes, and refine decisions across all sections of remanufacturing. ReManGPT can guide and support these expertise-related processes by combining knowledge from datasets and perception from real-time sensors. Through real-time natural language interaction, it can recommend which units are worth remanufacturing, propose an initial processing plan, and adjust these suggestions as new observations arise during operation. In later stages, it can retrieve historical records or design-related information to support reconditioning decisions. It can also provide structured guidance during reassembly when component upgrades or replacements are required. At the same time, the system can maintain a record of observations and decisions, thereby reducing the documentation workload on workers and enhancing the consistency and transparency of information throughout the remanufacturing workflow. 

\subsection{Electric Motors}

The increase in EV sales has also sharply increased the demand for traction electric motors. Beyond EVs, electric motors are widely used in various industries, including HVAC systems and most industrial equipment. The global market for electric motors was valued at \$197.78 billion in 2024 and is projected to reach \$322.08 billion by 2030 \cite{GrandViewResearch_ElectricMotorMarket_2024}. Many motors reach the recovery stage even when only minor components malfunction or overall efficiency decreases. As the market continues to grow, these early retirements are generating a rapidly expanding stock of recoverable motors. At the same time, recycling recovers less than 8\% of REEs in motors, although they account for more than 40\% of the total motor cost \cite{tiwari2021review}. This contrast underscores the strong potential of remanufacturing as a primary recovery solution for electric motors \cite{li2024circular}.

Electric motors are used widely across various industries and consumer applications, which introduces a large variance in size, structure, function, and value. These differences introduce significant uncertainty into the remanufacturing process, making it challenging to establish a consistent criterion for defining the EoL stages \cite{di2024circularity}. The current practice is still at the traditional manual stage that relies on human expertise to determine the proper remanufacturing process for each motor. This includes initial evaluation with limited product information, precise and complicated disassembly due to compact product design \cite{maani2024disassembly}, various reconditioning processes to meet remanufacturing objectives, and adaptive and high-precision reassembly operations \cite{tiwari2021review}. 

Compared with EV-LIBs and e-waste, current studies on electric motor remanufacturing remain relatively sparse and limited \cite{li2024circular}. Existing research primarily focuses on condition monitoring and vibration analysis during the usage phase \cite{magadan2020low}, which can provide valuable information for initial inspection and diagnostics. Detection models can mainly help separate motors from the solid waste stream, but offer little guidance for assessing their internal condition. Disassembly sequence planning methods also rely entirely on predefined knowledge for specific models and cannot generalize beyond those assumptions. All these methods struggle to meet the real-time requirements for providing detailed and context-aware instructions to human workers in a human-centric remanufacturing setting \cite{di2024circularity}.

ReManGPT can also be used to enhance the current human-centric electric motor remanufacturing process. It can retrieve relevant but scattered information, such as records, design files, digital passports, and prior cases, and combine these with real-time observations to perform initial inspections and evaluations. During disassembly, it can retrieve design data and technical specifications to generate real-time disassembly sequences and instructions, adjusting them as the internal structure becomes visible. This provides operators with clear and adaptive guidance that can improve both efficiency and process quality. In the reconditioning stage, it can synthesize updated inspection findings with operator feedback, which is expressed in natural language, to suggest an appropriate remanufacturing goal along with the corresponding reconditioning steps. Throughout the process, ReManGPT can also maintain a structured record of the motor and its components, which supports downstream reassembly. It can produce updated reassembly instructions that reflect the component history and prior operations. These capabilities reduce workload and mitigate the reliance on operator expertise. By strengthening information flow and decision support across stages, ReManGPT enhances the consistency and efficiency of current human-centric remanufacturing workflows, while supporting a gradual transition toward higher levels of automation. With improved manipulation capabilities, the decision assistant's outputs can be reformulated as instructions for robots or direct action commands, supporting a gradual transition toward a higher level of remanufacturing automation.

\section{Limitations and Challenges}
In this section, we explore the current challenges and limitations of LLMs in remanufacturing practices. 

\textbf{Hallucination:} Despite rapid progress in LLMs, hallucination remains a major barrier to their use in remanufacturing. It significantly compromises the trustworthiness of utilizing LLMs in remanufacturing practice \cite{rawte2023survey, xu2024hallucination}. In management-level decision support, LLMs may generate incorrect or impractical advice for complex scenarios. They may also fail to provide clear and reliable analyses of the data and reports. These outputs can reduce the trust of decision-makers and hinder the proper management of the entire framework, potentially reducing economic growth. More importantly, the hallucination in the technical operational process can lead to direct and severe failures. The LLMs might incorrectly detect and process the condition of the product, which may lead to an incorrect process design and remanufacturing schedule. This will not only lower efficiency and reduce overall economic output but also increase safety concerns, especially when considering remanufacturing in HRC and fully automated settings. An infeasible disassembly or assembly sequence may cause hazards and diminish the value of remanufactured parts. The improper inspection, sorting, and reconditioning processes, along with the instruction and troubleshooting guidance generated by the LLMs, will create new problems for the processes. The impacts of hallucination reduce confidence in deploying LLMs throughout the entire remanufacturing practice. 

To mitigate this issue, researchers are employing RAG to ground outputs in factual knowledge, integrating knowledge graphs for information construction, utilizing context engineering to reduce information loss during processing, and using LLM-based agents to enhance performance and reliability in complex scenarios. Despite these advancements, hallucination remains a significant and evolving challenge, particularly as LLMs integrate multi-modal inputs \cite{huang2025survey, zhang2025survey}.

\textbf{Interpretability and Trust:} Compared with many conventional deep learning models, LLMs can provide natural-language rationales and stepwise explanations for their outputs. However, the verifiability of these interpretations is a trust challenge for users in the remanufacturing domain. Although the stepwise explanations may make model outputs easier to follow, the inference process is often not fully transparent. Crucially, these explanations themselves can also suffer from hallucination, providing misleading information that does not reflect the models' actual decision logic. This lack of reliable interpretation directly impacts user trust and adoption. In the management framework, managers need transparent and traceable justifications based on verifiable data for strategic advice. In technical operations, experts need feasible and validated explanations for diagnostic and process suggestions. While the solution and explanation generated by LLMs seem comprehensive, they may not provide the validated and consistent information for experts. In that case, experts may not trust LLMs and tend to rely on other information or their own experience. This leads to poor user acceptance of integrating LLMs into practice, ultimately reducing the overall trustworthiness of using LLMs in remanufacturing operations. Further research into explainable AI (XAI) within the remanufacturing domain is vital to improve LLM transparency and build greater trust \cite{singh2024rethinking, cambria2024xai, dwivedi2023explainable}.

\textbf{Data Limitations:} The performance of LLMs relies heavily on the quality and quantity of the data. However, the LLMs in the remanufacturing domain face significant data limitations. There is a need to construct large, structured, comprehensive, and well-labeled datasets for different EoL products related to different aspects of the remanufacturing process. Moreover, the remanufacturing data is typically multi-modal at both the management and technical-operation levels. It includes long textual records, numerical logs, product images, videos, operational instructions, and sensor data. To implement LLMs in remanufacturing practice, it is challenging to effectively collect, annotate, and construct these heterogeneous data into an extensive and reliable dataset. The lack of datasets limits the capability to effectively train, fine-tune, augment, and evaluate the LLMs for specific remanufacturing tasks \cite{liu2024datasets}.

\textbf{Computational Resources:} The integration of LLMs into remanufacturing operations faces significant challenges due to their demanding computational resource requirements, which differ critically between the training and inference phases. For model training and fine-tuning, the demand for specialized hardware is exceptionally high. Developing or adapting LLMs for domain-specific tasks requires access to extensive GPU clusters and considerable computational power. This leads to high capital investment and energy consumption and creates significant financial barriers for many remanufacturing facilities. During the application of LLMs, the challenge shifts from extraordinary training loads to operational costs and real-time inference performance. Real-time applications in remanufacturing require LLMs to process an extensive amount of data, necessitating powerful and costly hardware for computation. LoRA and knowledge distillation \cite{xu2024survey} offer ways to deploy LLMs with limited resources and enhance the performance of small-scale LLMs; however, reducing overall computational requirements for practical deployment in remanufacturing is still challenging \cite{bai2024beyond}.

\textbf{Data Privacy and Security:} Another significant challenge for integrating LLMs into remanufacturing practice is data privacy and security. These data include essential product design, process parameters, production records, financial information, and personal details. Because local deployment of LLMs is computationally expensive, many practices rely on cloud-based services such as ChatGPT APIs, which increases the risk of data exposure to third parties and raises legal and compliance concerns \cite{yao2024survey}. Also, when training or input data are contaminated, LLMs may produce unsafe outputs that lead to operational hazards. This highlights the ongoing challenge of ensuring that the data used by LLMs is secure and trustworthy \cite{das2025security}.

\section{Future Directions}

\subsection{Large Language Models}
This section outlines potential future research directions for integrating LLMs into remanufacturing. These directions aim to address current deployment limitations and support more reliable intelligent remanufacturing systems.

To mitigate hallucinations in remanufacturing, LLMs should retrieve task-relevant information through Retrieval-Augmented Generation (RAG) before generating responses. This enhances model reliability by linking responses to verified remanufacturing records and operational instructions. The remanufacturing knowledge graph (KG) can be used within the RAG pipeline to organize information and improve generation accuracy. The scope and relations in the KG should be tailored to the specific remanufacturing tasks instead of general purpose applications. At the generation stage, the multi-agent setups can integrate remanufacturing-specific agent skills to further reduce hallucinations. Each response generated by LLMs should follow a defined workflow to meet specific requirements for various tasks across the remanufacturing process. For deployment, companies need to prepare and maintain datasets and KGs for RAG purposes. They also need to develop agent skills metadata to meet requirements for their specific remanufacturing processes. Additionally, context engineering is crucial in remanufacturing, where large amounts of information are frequently updated. By retaining key case information and relevant interaction history over long-horizon tasks, context engineering helps reduce hallucinations caused by missing or outdated context.

While hallucination mitigation enhances the factual reliability of LLM-generated information, future research needs to improve interpretability and trust to ensure the safe and transparent deployment of LLMs in remanufacturing systems. One key direction could also be to integrate the KGs that are designed for remanufacturing. Outputs should be linked to verified sources, specific constraints, and different decision rules. This allows users to trace and understand the reason for the generated results. XAI methods also need to be incorporated to provide explanations for the results to improve the interpretability. These two methods together can improve the user trust while deploying LLMs in remanufacturing practice.

Addressing data availability is a key direction for future research to enable LLM integration in remanufacturing practice. These data are essential for both model development and deployment in remanufacturing settings. Future work needs to develop efficient methods to construct high-quality remanufacturing domain-specific datasets from industrial remanufacturing categories and processes, including collection, annotation, construction, and information updating. The datasets should contain multi-modal aligned information, including production logs, quality reports, product images, operational instructions, and sensor data. This multi-modal dataset requirement can support training, fine-tuning, and evaluation for both text-based and multi-modal LLMs. It can also support LLM-based applications by providing reliable information for RAG pipelines, remanufacturing KGs, and agent skills.

Regarding the high computational resource constraints for LLMs in remanufacturing, research should focus on improving the efficiency and deployability of LLMs. One direction is utilizing lightweight model architectures with optimization methods to maintain performance with limited computational costs. Knowledge distillation can transfer capabilities from large-scale LLMs to smaller models \cite{xu2024survey}. When integrated with reinforcement learning-based tuning using high-quality data, distilled models can achieve comparable or improved performance with limited computational resources \cite{guo2025deepseek}. Similarly, the mixture-of-experts (MoE) architecture can also be an effective tool to reduce the computational cost while maintaining the capacity of large-scale LLMs for complex reasoning tasks \cite{liu2024deepseek}. Parameter-efficient fine-tuning methods, such as LoRA, enable task- and domain-specific adaptation of LLMs on limited hardware while maintaining model performance. All these methods present practical pathways for deploying LLMs in remanufacturing practice. While the previously mentioned methods require modifying the internal parameters of LLMs to achieve domain-specific performance, prompt engineering offers an efficient approach. Well-designed prompts can leverage the in-context learning ability of pre-trained models to perform complex reasoning without altering the models themselves, thereby reducing the computational cost for model training or fine-tuning during deployment.

Ensuring data privacy and security across the entire LLM lifecycle is critical for integrating LLMs into remanufacturing. Future research should establish practical protocols and regulatory frameworks to monitor how sensitive remanufacturing-related data is collected, stored, and used by LLM-based systems. Differential privacy \cite{charles2024fine} and federated learning \cite{kuang2024federatedscope} should be adopted to prevent raw industrial data from being exposed during model training and deployment stages. At the same time, mechanisms such as LLM firewalls can help prevent unauthorized access, data poisoning, and prompt injection, ensuring interaction security between LLMs and shop-floor remanufacturing operational systems. Continuous monitoring of data use and model behavior is crucial for maintaining transparency and trust in remanufacturing LLM applications. Addressing ethical bias in training data is another integral part of data security. Reducing biased information in datasets can enhance the reliability of model outputs and prevent the use of unexpected or unsafe information that could affect decision-making during remanufacturing operations \cite{rathod2025privacy}.

\subsection{Vision Language Action Models}

In this section, we further explore the role of vision-language-action (VLA) models as a possible robotic-execution module within the ReManGPT framework for future intelligent remanufacturing. VLA models have attracted growing attention because they can map perception and language inputs directly to robot actions. However, the preliminary case study shows that current VLA models remain insufficient for complex remanufacturing environments. The following discussion summarizes the main limitations and possible research directions.

Because VLA data collection and training are costly, most researchers fine-tune pre-trained VLA models for domain-specific adaptation. While offering various fine-tuning options for researchers, these models are also limited by input modality and quality constraints, including camera coverage, robot state information, image resolution, and language instruction quality. The low resolution and limited number of cameras may lower the confidence of VLA models in determining the current states, especially in the remanufacturing domain for small-scale objects with high operational precision requirements. VLA models may struggle to localize the target in indistinct images, and the robot may also completely block the camera view. The robot state information is not considered in some VLA models, even though it could be an important modality for long-horizon remanufacturing tasks. Robot state information may be needed to indicate the completed steps, current configuration, and next feasible action. In a complex environment with limited space, the vision input may not always be sufficient. Depth information can enhance identification and localization accuracy, and force and tactile sensor information can improve the robot's manipulation, enabling it to complete the entire operation in a contact-rich setting. These limitations suggest that remanufacturing-oriented VLA models may need richer multi-modal inputs during pre-training or domain adaptation. The case study also suggests that language input remains underused in current VLA settings. The current language is simple and consistent throughout the task, such as ``Remove the RAM modules'', without providing additional instructions, such as step-by-step decompositions, the current environment, tool information, or historical operations, to guide the robot's action. We believe that, as in LLM-generated instructions for human operators, language should play a much more important role in VLA models \cite{liu2025vision}. 

Another limitation of applying VLA models to a complex environment is the distribution shift between the training and real-world experiments. Although closed-loop action prediction can reduce some open-loop error accumulation, execution failures may still compound when the robot enters states that are underrepresented or absent in the training data. Moreover, the massive pre-training dataset does not contain complex environmental settings, such as disassembly, making the generalization capability highly dependent on the data collected specifically for our setting. Consequently, a substantial amount of high-quality data for pre-training or fine-tuning purposes is crucial for utilizing VLA models in the remanufacturing domain. A shared autonomy control strategy can be employed for high-quality data collection, thereby reducing noise from teleoperation demonstrations \cite{liu2026self}. In this setup, the motion before manipulation is controlled by a human, and the manipulation operation is executed under a pre-defined program to guarantee consistency and precision, and enhance the dexterity of the operation. Deploying a simulation environment could be an efficient and essential way to improve the quality and increase the amount of collected data. With the recent achievements in 3D Gaussian splatting and high-quality 3D reconstruction \cite{kerbl20233d}, the gap between the real world and simulation has been significantly narrowed. The EoL product can be reconstructed directly and efficiently within the simulation world with quality comparable to that of the real product. The robot can move to any position inside the simulation to collect diverse observations and perform manipulation with realistic physical interactions. Since the simulation data has high similarity with the real-world data, it significantly enhances the sim-to-real transformation. Any trained skill that the robot learned from the simulation world can be more directly deployed in the real world. 

Utilizing reinforcement learning (RL) methods in conjunction with the VLA model offers an alternative approach to addressing the limitations mentioned above, particularly the poor real-world performance issue resulting from data scarcity and distribution shift. RL can be applied offline or online to adapt a pre-trained VLA to a new environment for various subtasks. Although RL increases reliability, the training time for each subtask is not ideal for complex remanufacturing operations \cite{lu2025vla}. Alternatively, similar to RLHF for LLMs, an RL framework in which VLA models can learn from human experts provides an exciting and feasible approach. Humans can correct the action of the robot during execution. These corrections are recorded as high-quality feedback data for further RL fine-tuning to enhance performance and mitigate the data limitations. While the framework has demonstrated strong performance on daily-life tasks \cite{amin2025pi}, its applicability to complex remanufacturing has yet to be validated.

\section{Conclusion}

This paper provides a forward-looking review and analysis of the role of LLMs in remanufacturing automation, grounded in a brief critical review of existing LLM-related studies relevant to remanufacturing. The review shows that these methods are beginning to support work across distinct remanufacturing stages, while also extending to higher-level design and process functions and to more integrated cross-stage capabilities for decision-making, planning, and execution. However, uncertainty remains the core challenge in remanufacturing, and current research has not yet resolved it robustly across stages. Higher levels of automation have therefore not yet been realized. The proposed ReManGPT framework should serve to integrate LLM-related capabilities in a modular way and apply them to different remanufacturing tasks as needed, rather than being taken as a completed solution. From this perspective, the main findings of this paper can be discussed through four closely related questions.

The first question is: \textbf{How can reliance on experienced workers for knowledge retrieval and decision-making in remanufacturing be reduced?} This issue persists because most of the information that is required to support remanufacturing is fragmented and multi-modal, including user manuals, historical records, process documents, and prior operational experience. Operators often need considerable effort to locate and understand relevant information, and then to decide how to properly handle EoL products under various conditions. In many cases, they still rely on experienced workers for help. From this perspective, a key contribution of LLM-related methods is their ability to retrieve scattered multi-modal information, interpret it in natural language, and transform it into usable guidance for operators. This is particularly valuable in knowledge-intensive remanufacturing tasks, where workers often need to make case-specific decisions during operation. In this context, ReManGPT leverages LLM capabilities to reduce reliance on experienced workers and improve decision-making in practical remanufacturing tasks. The repair-assistance case study illustrates this role through structured diagnostic guidance and stepwise instruction generation. However, such reliance has not been fundamentally reduced. In practice, remanufacturing knowledge remains difficult to collect, structure, and maintain, particularly under operational and privacy constraints. Without sufficient transparency in the process and validation of the outputs, operators may still hesitate to trust system-generated guidance.

The second question is: \textbf{How does one develop a unified, general-purpose model to dynamically support decision-making, planning, and robot control to handle uncertainty in remanufacturing operations?} Current research has not yet resolved this problem, as most existing methods remain task-specific and are developed for limited settings. ReManGPT also does not provide such a model or a complete solution in this sense. Its contribution lies in offering a modular framework that integrates diverse capabilities and applies them to various remanufacturing tasks as needed. In this way, it provides a more systematic basis for handling uncertainty across stages. This is illustrated by the disassembly sequence planning case study, in which ReManGPT combines multi-modal perception, knowledge retrieval, and reasoning to identify high-value target components, infer blocked components, and generate a feasible disassembly sequence with operational and tool requirements under uncertain product conditions. The VLA-based robotic disassembly case further shows that the same framework can be extended directly to robotic action generation, although this remains preliminary. This problem therefore remains open.

The third question is: \textbf{How can the cognitive gap be bridged so that operators understand the reasons and procedures for their tasks, while also being able to adjust robotic plans based on real-time observations without programming expertise?} In remanufacturing, operators may still be unable to understand the rationale behind system outputs or translate new observations into corresponding robotic execution. Current LLM-related methods have improved the interpretability of task logic and generated instructions, making it easy for operators to understand and follow the result. ReManGPT makes this role more explicit at the framework level and illustrates it in the repair-assistance case. The direct transition from observation to robotic execution remains less developed. Both the review and the ReManGPT framework suggest that operator observations and revised requirements can be carried further toward executable robotic actions, but this has not yet been established as a reliable capability for non-programming operators in practice. The cognitive gap has therefore been reduced, but not closed.

The last question is: \textbf{How can the automation level be elevated from task-specific execution to generalized automation while effectively minimizing human intervention in the current remanufacturing loop?} The current work has not yet reached this level of automation. Most existing methods remain task-specific and leave the decision-making and process adjustment to human expertise. Section 2.6 shows that LLM-related methods are beginning to support real-time planning, robotic control, and end-to-end action generation, suggesting a possible path beyond task-specific execution. ReManGPT takes this one step further by organizing these capabilities within the remanufacturing workflow, combining perception, reasoning, planning, and execution. This direction is reflected in the VLA-based robotic disassembly case, where robotic actions are generated within the framework. However, the preliminary results suggest that the current systems cannot complete a full remanufacturing task in real practice. Generalized automation with reduced human intervention has therefore not yet been achieved.

The question for LLMs in remanufacturing is no longer whether they can contribute, but how their capabilities can be used under real uncertainty and extended beyond isolated tasks. While progress is evident, current research remains uneven and largely task-specific. ReManGPT is better understood as a framework for applying these capabilities in practice than as a complete solution. Progress is further constrained by the scarcity of remanufacturing data, which limits validation and broader deployment. Intelligent remanufacturing with robust cross-stage support and substantially reduced human intervention has yet to be realized.

\bibliographystyle{IEEEtran}
\bibliography{ref}{}

\end{document}